\documentclass[journal]{IEEEtran}
\IEEEoverridecommandlockouts
\usepackage{cite}
\usepackage{amssymb,amsfonts,amsthm,amsmath}
\usepackage{algorithm}
\usepackage{algorithmic}
\usepackage{graphicx}
\usepackage{textcomp}
\usepackage{xcolor}
\usepackage{comment}
\usepackage{epstopdf}
\usepackage{subfigure} 
\newtheorem{remark}{Remark}

\def\BibTeX{{\rm B\kern-.05em{\sc i\kern-.025em b}\kern-.08em
    T\kern-.1667em\lower.7ex\hbox{E}\kern-.125emX}}
\begin{document}

\title{Delay-Doppler Domain Signal Processing Aided OFDM (DD-a-OFDM) for 6G and Beyond}
        
\author{Yiyan~Ma,~\IEEEmembership{Member,~IEEE,}
        Bo~Ai*,~\IEEEmembership{Fellow,~IEEE,}
        Jinhong~Yuan,~\IEEEmembership{Fellow,~IEEE,}
        Shuangyang~Li,~\IEEEmembership{Member,~IEEE,}
        Qingqing~Cheng,~\IEEEmembership{Member,~IEEE,}
        Zhenguo~Shi,~\IEEEmembership{Member,~IEEE,}
        Weijie~Yuan,~\IEEEmembership{Senior Member,~IEEE,}
        Zhiqiang~Wei,~\IEEEmembership{Member,~IEEE,}
        Fan~Liu,~\IEEEmembership{Senior Member,~IEEE,}
        Akram~Shafie,~\IEEEmembership{Member,~IEEE,}
        Guoyu~Ma,~\IEEEmembership{Member,~IEEE,}
        Yunlong~Lu,~\IEEEmembership{Member,~IEEE,}
        Mi~Yang,~\IEEEmembership{Member,~IEEE,}
        ~and~Zhangdui~Zhong,~\IEEEmembership{Fellow,~IEEE}
        
\thanks{Y. Ma, B. Ai, G. Ma, M. Yang, Y. Lu, and Z. Zhong are with the Beijing Jiaotong University. J. Yuan and A. Shafie are with the University of New South Wales. S. Li is with the Technical University of Berlin. Q. Cheng and Z. Shi are with the Queensland University of Technology. W. Yuan is with the Southern University of Science and Technology. Z. Wei is with the Xi'an Jiaotong University. F. Liu is with the Southeast University. (Corresponding authors: boai@bjtu.edu.cn).}

}
\maketitle

\begin{abstract}
High-mobility scenarios will be a critical part of 6G systems. Since the widely deployed orthogonal frequency division multiplexing (OFDM) waveform suffers from subcarrier orthogonality loss under severe Doppler spread, delay-Doppler domain multi-carrier (DDMC) modulation systems, such as orthogonal time frequency space (OTFS), have been extensively studied. While OTFS can exploit time-frequency (TF) domain channel diversity, it faces challenges including high receiver complexity and inflexible TF resource allocation, making OFDM still the most promising waveform for 6G. In this article, we propose a DD domain signal processing-aided OFDM (DD-a-OFDM) scheme to enhance OFDM performance based on DDMC research insights. First, we design a DD-a-OFDM system structure, retaining the conventional OFDM transceiver while incorporating DD domain channel estimation and TF domain equalization. Second, we detail DD domain channel estimation using discrete TF pilots and prove that TF domain inter-carrier interference (ICI) could be transformed into DD domain Gaussian interference. Third, we derive closed-form Cram\'{e}r-Rao lower bounds (CRLBs) for DD domain channel estimation. Fourth, we develop the maximum likelihood (ML) estimator, along with a corresponding TF domain equalizer. Numerical results verify the proposed design, showing that DD-a-OFDM reduces the bit-error rate (BER) compared to conventional OFDM and {\color{black}can achieve the same CRLB performance as the classical embedded-pilot OTFS}.
\end{abstract}


\section{Introduction}

In the 6G era, high-speed mobile scenarios will become increasingly prevalent compared to 5G and previous systems, including intelligent railway systems \cite{ailaoshi,ailaoshi2}, low-altitude wireless networks \cite{WEIJIE,jingli}, and autonomous vehicular networks \cite{ziji2,hbx,hbx2}, etc. In these high-mobility scenarios, the performance of {\it conventional} orthogonal frequency division multiplexing (OFDM) systems becomes significantly limited \cite{Andrews}. The combined effects of high mobility and multipath propagation induce Doppler spread \cite{xuejian1,xuejian2,zhengyu1,zhengyu2}, which not only transforms the wireless channel into a linear time-varying (LTV) system but also destroys subcarrier orthogonality, resulting in an error floor. Consequently, waveform design for high-mobility scenarios has remained a consistently active research topic. \par

From 2017 to present, delay-Doppler (DD) domain multicarrier modulation (DDMC) waveforms have undergone systematic investigation. Representative DDMC schemes include orthogonal time frequency space (OTFS) modulation \cite{hadani,zhiqiangmeg,xuehan1}, orthogonal delay-Doppler division multiplexing (ODDM) \cite{lin,zijitcom}, Zak-OTFS \cite{zakotfs2,zakotfs,zakotfs3}, etc. Unlike OFDM, DDMC waveforms design transceivers by leveraging the quasi-static nature of the DD domain wireless channel, where sophisticated DD domain signal processing strategies have been developed. As a result, DDMC schemes can achieve full time-frequency (TF) domain diversity and demonstrate superior reliability to OFDM in high-mobility environments \cite{Raviteja}.\par

Despite the promise, DDMC systems present two fundamental challenges. Firstly, DD domain multiplexing inherently limits the flexibility of TF domain resource allocation \cite{Andrews}. Secondly, the extended DD domain data frames increase the complexity of conventional linear equalizers quadratically compared to OFDM systems \cite{zijitcom}. Given OFDM's mature deployment infrastructure and the practical challenges of replacing it entirely in high-mobility scenarios, OFDM remains the most promising waveform candidate for 6G systems \cite{Andrews}. Given the critical juncture of 6G development, OFDM limitations, and DDMC advancements, we propose enhancing OFDM's reliability in high-mobility scenarios through DD domain signal processing, leading to our proposed DD domain signal processing-aided OFDM (DD-a-OFDM) framework.\par

Classical research has explored OFDM performance enhancement through DD domain signal processing. The basis expansion model (BEM) represents a notable approach, modeling complex exponential phase shifts from resolvable Doppler frequencies as basis functions, with time-invariant DD domain channel coefficients as BEM factors \cite{BEM}. This allows time-domain channel fading representation via BEM factors and basis functions. While BEM reduces channel matrix dimensionality and enables low-complexity OFDM receivers, its fixed basis functions cannot adapt to DD channel sparsity variations, and per-symbol processing limits achievable DD domain diversity gain. Recently, the coexistence of OTFS and OFDM has been investigated \cite{akram}, where OTFS principles are applied over OFDM infrastructure assisted by multiple cyclic prefixes (CPs). However, this framework requires DD domain data embedding, channel estimation, and equalization, and faces similar challenges as standalone DDMC systems.

{\color{black}Very recently, DD domain channel estimation using TF domain discrete pilots has been investigated in \cite{gong} and \cite{R1_Shaw}, demonstrating the potential for reliable OFDM in high-mobility scenarios without altering the transmit waveform. Furthermore, \cite{R2_Marchese} proposed a robust OFDM framework utilizing DD domain superimposed pilots to enhance resource efficiency. {\color{black}However, \cite{gong} and \cite{R1_Shaw} primarily emphasize algorithmic gains but lack rigorous theoretical analyses of estimation limits, especially under fractional Doppler and delay leakages. Additionally, the superimposed pilot design in \cite{R2_Marchese} could introduce pilot-data interference and requires an advanced receiver.}} Our preliminary work \cite{wlc} also initiated DD-based designs, yet overlooked the impact of inter-carrier interference (ICI) and rigorous derivations. Consequently, establishing a unified theoretical framework and implementation for DD-a-OFDM under severe ICI remains an open challenge.\par

Recognizing these research gaps, in this work, we propose a DD-a-OFDM system with novel receiver designs, along with in-depth theoretical analysis. {\color{black}The main contributions of this work are summarized as follows:}

\begin{itemize}
\item We propose a DD-a-OFDM architecture that harmonizes the conventional TF domain multiplexing with DD domain channel estimation. While maintaining the TF domain structure to ensure seamless backward compatibility with 5G NR deployments, the proposed system exploits the DD domain for high-precision multipath parameter estimation. This approach could enhance robustness against doubly-selective fading compared to traditional TF domain estimation methods, which are often limited by {\color{black}TF domain} channel observations.

\item We establish a comprehensive analytical framework for DD domain channel estimation within the DD-a-OFDM system. Specifically, we formulate the estimation model using TF domain discrete pilots and analytically demonstrate that the TF domain ICI is transformed into statistically tractable Gaussian noise in the DD domain. We further derive the Cram\'{e}r-Rao Lower Bounds (CRLB) to quantify the theoretical limits of estimation accuracy. Based on these insights, a DD domain maximum likelihood (ML) estimator and a corresponding minimum mean square error (MMSE) equalizer are developed.

\item Extensive simulations are conducted to evaluate the mean square error (MSE) and the bit error rate (BER) performance. First, the correctness of the derived CRLB is verified by {\color{black}the} MSE of the ML estimator. Besides, the results demonstrate that DD-a-OFDM significantly outperforms conventional OFDM due to its superior ICI mitigation capabilities. {\color{black}Moreover, compared to the classical embedded-pilot OTFS, the proposed DD-a-OFDM guarantees equivalent theoretical CRLB performance under identical pilot overhead. Furthermore, by effectively transforming ICI into a Gaussian process, DD-a-OFDM circumvents the severe non-Gaussian data-to-pilot leakage inherent in OTFS Dirichlet sampling.}
    
\end{itemize}

{\color{black}The remainder of this article is organized as follows.} Section II presents the DD-a-OFDM system model. Section III details discrete TF pilot-based DD domain channel estimation principles. Section IV describes DD domain channel estimators and TF domain equalizer designs. Section V provides simulation results and analysis. Finally, Section VI concludes the work and outlines future research directions.

\section{System Model of DD-a-OFDM}

This section first introduces LTV channel representations across different domains. Building upon the DD domain characteristics of LTV channels, we then present the transmission framework for DD-a-OFDM. Finally, we briefly review the input/output (I/O) relationship of conventional OFDM systems.
\subsection{LTV Channels Representations}
\label{LTV}
The LTV channels can be described in different domains. In the traditional communication systems, the channel impulse response (CIR) is useful for system design, which models the LTV channel in the time-delay (TD) domain and is written as:
\begin{equation} 
h_{\rm TD}(t,\tau) = \sum^{P}_{i=1} h_i e^{j2\pi \nu_i t} \delta(\tau-\tau_i), 
\label{CIR} 
\end{equation}
where $\tau_i \in [0, \!\tau_{\max}]$, $\nu_i \!\in\! [-\nu_{\rm max}, \nu_{\rm max}]$, and $h_i$ denote the delay, Doppler, and fading coefficient of the $i$-th path, respectively, $P$ represents the number of multipath, and $\tau_{\rm max}$ and $\nu_{\rm max}$ are maximum delay and Doppler shift, respectively.\par

Besides, in the conventional OFDM systems, the wireless channel in the TF domain, i.e., the channel transfer function (CTF), is usually concerned. The CTF characterizes the channel TF selectivity, which can be obtained by applying Fourier transform on the CIR in (\ref{CIR}) along the delay domain as: 
\begin{equation}
h_{\rm TF}(t,f) = \sum^{P}_{i=1} h_i e^{j2\pi \nu_i t} e^{-j2\pi \tau_i f}.
\label{CTF}
\end{equation}

In the DD domain, the LTV channel is modeled by the channel spreading function (CSF). The CSF describes the spreading of the transmit signal in the TF domain, i.e., the delay spread and Doppler spread. By applying Fourier transform along time domain and inverse Fourier transform along frequency domain on the CTF in (\ref{CTF}), the CSF is obtained by \cite{matz} :
\begin{equation} 
h_{\rm DD}(\tau,\nu) = \sum^{P}_{i=1} h_i \delta(\tau-\tau_i)\delta(\nu-\nu_i).
\label{CSF} 
\end{equation}

Among the above channel characterizations, the LTV channel can be regarded as time-invariant in different time and frequency scales. Specifically, the CSF can be regarded as quasi-invariant when there are no notable multipath birth-and-death events, i.e., within the stationary region \cite{ziji3}. Meanwhile, the CTF can be regarded as invariant within the channel coherence region \cite{DTse}. More importantly, literature shows that the channel stationary region could be larger than the coherence region by one order of magnitude \cite{ziji4}. Therefore, estimating the DD domain CSF will be less challenging than estimating the TF domain CTF, since the CSF varies more slowly, which motivates the design of DD-a-OFDM.\par
\subsection{Transmission Framework}
\begin{figure}
  \centering
  \includegraphics[width=0.48\textwidth]{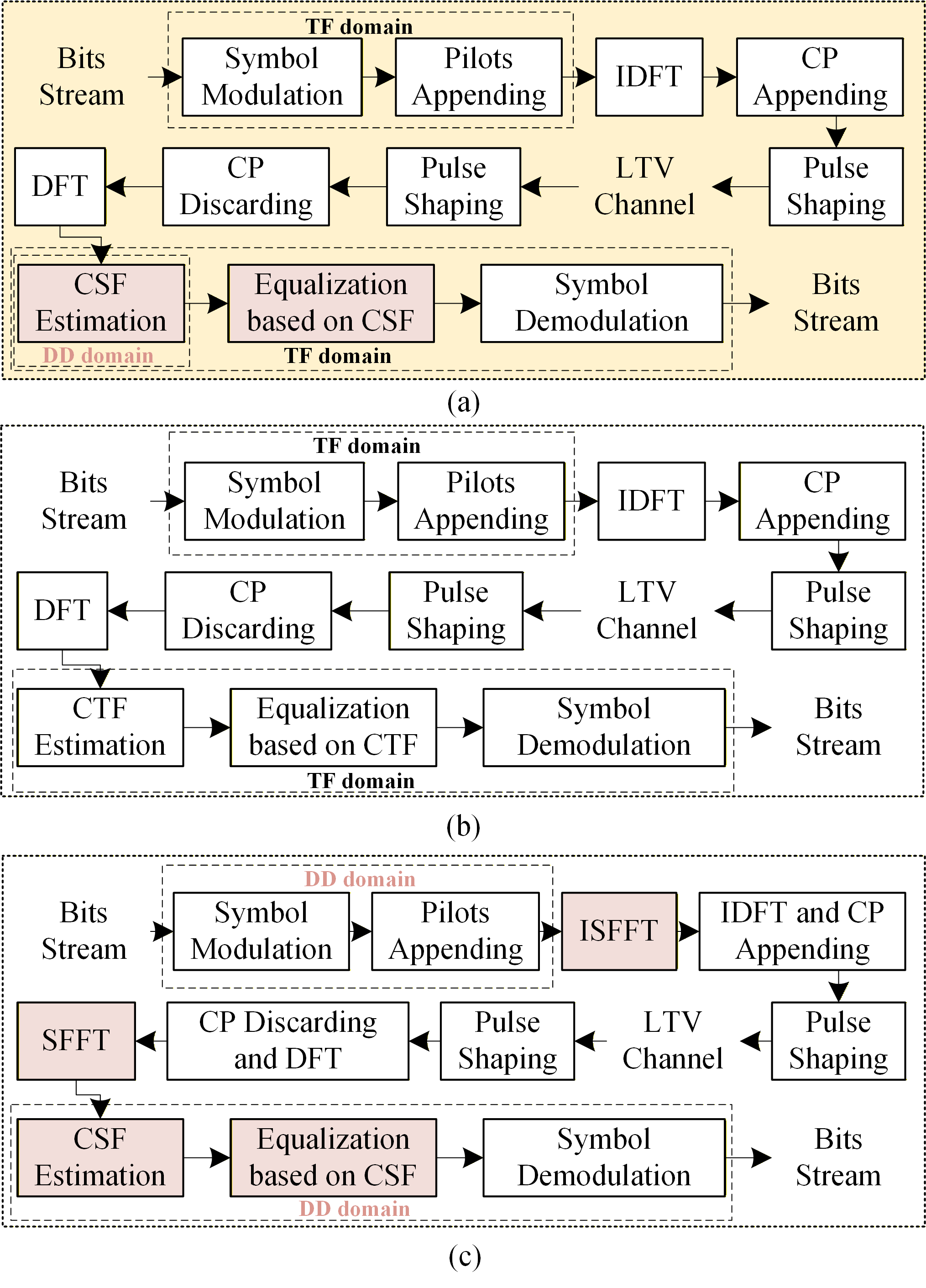}
  \caption{Transmission Diagrams of: (a) Proposed DD-a-OFDM, (b) {\color{black}Conventional} OFDM, and (c) OTFS. {\color{black}The modules that differ from those in} the conventional OFDM systems are highlighted in red.}\label{FIG1}
    \vspace{-1em}
\end{figure}

The DD-a-OFDM system is designed based on the characteristics of CSF and OFDM framework. For illustration, the diagram of the DD-a-OFDM system is given in Fig. \ref{FIG1}(a), compared to that of the conventional OFDM and OTFS\footnote{\color{black}The OTFS structure in Fig. 1(c) represents the original ISFFT-OTFS \cite{hadani}. Without loss of generality, it can be equivalently represented or replaced by the Zak-OTFS implementation \cite{zakotfs3}. Note that while the cascaded ISFFT and IDFT partially cancel out, we retain this representation to clearly distinguish the Doppler-domain IDFT in OTFS from the frequency-domain IDFT in OFDM.}.\par

As illustrated in Fig. \ref{FIG1}, the proposed DD-a-OFDM system retains the fundamental architecture of conventional OFDM, maintaining its TF domain multiplexing structure, inverse discrete Fourier transform (IDFT) and discrete Fourier transform (DFT), cyclic prefix (CP) handling, and pulse shaping mechanisms. The key innovation lies in replacing conventional CTF-based processing with novel CSF-oriented channel estimation and equalization techniques. Unlike OTFS systems that implement full DD domain multiplexing, DD-a-OFDM strategically applies DD domain signal processing while preserving OFDM's native framework. This hybrid design makes DD-a-OFDM possible to combine the respective strengths of both paradigms, i.e., OFDM's TF domain multiplexing and OTFS's superior high-mobility robustness. 

\subsection{I/O Relationship of OFDM Systems under High-Mobility}
Since the framework of DD-a-OFDM is {\color{black}the} same as conventional OFDM except for channel estimators and equalizers, the I/O relationship of the conventional OFDM under high mobility is briefly introduced for designs in subsequent sections.\par

Denote the number of OFDM symbols and subcarriers as $N$ and $M$, the time duration of an OFDM symbol and the subcarrier interval as $T$ and $\Delta f$, respectively. Let ${\bf x}_{\rm TF} \in {\mathbb C}^{M \times N}$ represent the transmitted symbols in the TF domain, with $m \in \{0,1,...,M-1\}$, $n \in \{0,1,...,N-1\}$. After IDFT and CP appending, the TD domain discrete symbols ${\bf x}_{\rm TD} \in {\mathbb C}^{(M+L_{\rm CP}) \times N}$ are given by:
\begin{equation}
    {\bf x}_{\rm TD}[l,n] = \frac{1}{\sqrt{M}} \sum_{m=0}^{M-1} {\bf x}_{\rm TF}[m,n] e^{j\frac{2\pi ml}{M}},
    \label{x_td}
\end{equation}
where $l \in \{-L_{\rm CP},\ldots,M-1\}$ is the index of delay domain symbols with interval $T_s = \frac{T}{M}$, $L_{\rm CP}$ refers to the length of CP symbols, and the CP duration is $T_{\rm CP}$ with $T_{\rm CP}\ge \tau_{\rm max}$. Besides, the duration of an OFDM symbol with CP is denoted as $T_{\rm sym} = T+T_{\rm CP} = (M + L_{\rm CP})T_s$. Afterwards, the discrete TD domain symbols are pulse-shaped by $g_{\rm tx}(t)$ as:
\begin{equation}
    s(t) = \sum_{n=0}^{N-1} \sum_{l=-L_{\rm CP}}^{M-1} {\bf x}_{\rm TD}[l,n] g_{\rm tx}(t - lT_s - nT_{\rm sym}).
    \label{s_t}
\end{equation}

After transmitting $s(t)$ over the LTV channel defined in Section \ref{LTV} and ignoring noise for brevity, the received time domain signal is given as:
\begin{equation}
r(t)=\sum_{i=1}^{P} h_i s\left(t-\tau_i\right) e^{j 2 \pi \nu_i t}.
\end{equation}

 At the receiver, $r(t)$ is firstly matched filtered by $g_{\rm rx}(t)$ as:
\begin{equation}
\begin{aligned}
y(t) & = \sum_{i=1}^{P} h_i \int e^{j2\pi \nu_i \tau} s(\tau - \tau_i) g_{\rm rx}(t - \tau) d\tau.
\end{aligned}
\label{y_t}
\end{equation}

By substituting $s(t)$ in (\ref{s_t}) into (\ref{y_t}), $y(t)$ turns into:
\begin{equation}
\begin{aligned}
    y(t) &= \sum_{i=1}^{P} h_i \sum_{n=0}^{N-1} \sum_{l=-L_{\rm CP}}^{M-1}{\bf x}_{\rm TD}[l,n] e^{j2\pi \nu_i (nT_{\rm sym} + lT_s + \tau_i)} \\
    &\quad \times A_{g_{\rm tx}, g_{\rm rx}}(t - nT_{\rm sym} - lT_s - \tau_i, \nu_i),
    \end{aligned}
\end{equation}
where $A_{g_{\rm tx}, g_{\rm rx}}(\tau,\nu)$ is the ambiguity function of transceiver pulses, which is given by:
\begin{equation}
    A_{g_{\rm tx}, g_{\rm rx}}(\tau,\nu) = \int g_{\rm tx}(u) g_{\rm rx}(\tau - u) e^{j2\pi \nu u} du.
\end{equation}

Then, the receiver will sample $y(t)$ at $t = n'T_{\rm sym} + l'T_s$ to acquire the discrete TD domain symbols, where $n' \in \{0,1,...,N-1\}$, $l' \in \{-L_{\rm CP},-L_{\rm CP}+1,...,0,1,...,M-1\}$:
\begin{equation}
    \begin{aligned}
    {\bf y}_{\rm TD}[l',n'] &\!=\! \sum_{i=1}^{P} h_i\!\sum_{n=0}^{N-1} \sum_{l=-L_{\rm CP}}^{M-1}{\bf x}_{\rm TD}[l,n] e^{j2\pi \nu_i (nT_{\rm sym} \!+ lT_s \!+ \tau_i)} \\
             &\quad \times A_{g_{\rm tx}, g_{\rm rx}}((n'-n)T_{\rm sym} + (l'-l)T_s - \tau_i, \nu_i).
    \end{aligned}
    \label{y_td1}
\end{equation}

We note that when $T_s$ is quite short, $A_{g_{\rm tx}, g_{\rm rx}}(\tau,\nu)$ can be approximated by $e^{j2\pi \nu \tau}\int g_{\rm tx}(u) g_{\rm rx}(\tau - u) du$. Furthermore, it is possible to design $g_{\rm tx}(t)$ and $g_{\rm rx}(t)$ to ensure $\int g_{\rm tx}(u) g_{\rm rx}(\tau - u) du \approx 0$ for $|t| > T_s/2$, e.g., they are root-raised-cosine pulses. By further considering that $\tau_i$ is on-grid with respect to (w.r.t.) the delay resolution $\frac{1}{M\Delta f}$, i.e., $l_i = \tau_i M\Delta f$ is an integer, {we observe that $A_{g_{\rm tx}, g_{\rm rx}}((n'-n)T_{\rm sym} + (l'-l)T_s - \tau_i, \nu_i)$ approaches unity when $n = n'$ and $l = l' - l_i$ due to matched filtering, and becomes negligible otherwise.} Thus, (\ref{y_td1}) can be simplified into\footnote{{\color{black}When $\tau_i$ is off-grid or practical pulse shapes are considered, the channel energy disperses across multiple delay bins in the DD domain, which translates to subcarrier-specific phase rotations and amplitude scalings in the TF domain. Consequently, while these effects require a refined signal model for precise DD-domain channel estimation, they strictly preserve the statistical properties of the i.i.d. data-induced interference that will be derived in (\ref{I_DD}). Thus, the derivation principles of the CRLB via FIM partial derivatives in Section III-E and the ML estimator in Section IV-A remain applicable. For simplicity, we focus on on-grid delay scenarios with rectangular pulses, while off-grid delay extensions will be addressed in future research.}}:
\begin{equation}
     {\bf y}_{\rm TD}[l',n'] \approx \sum_{i=1}^{P} h_i e^{j2\pi \nu_i (n'T_{\rm sym} + l'T_s)} {\bf x}_{\rm TD}[l' - l_i,n'].
     \label{y_td2}
\end{equation}
Finally, after discarding CP and performing DFT, the received symbols in TF domain can be obtained:
\begin{equation}
    {\bf y}_{\rm TF}[m',n'] = \frac{1}{\sqrt{M}}\sum_{l=0}^{M-1} {\bf y}_{\rm TD}[l',n'] e^{-j\frac{2\pi m'l}{M}}.
    \label{y_tf}
\end{equation}


Based on (\ref{x_td}), (\ref{y_td2}), and (\ref{y_tf}), we can obtain the matrix form of the TF domain input-output relationship of OFDM systems under the LTV channel as:
\begin{equation}
{\bf y}_{\rm TF}^n = \underbrace{\sum^{P}_{i = 1} h_i {\bf F}{\bf \Pi}^{l_i} {\bf \Lambda}_{n}^{(k_i)} {\bf F}^{H}}_{{\bf H}_{\rm TF}^n}{\bf x}_{\rm TF}^n + {\bf w},
\label{OFDMIOTD}
\end{equation}
where ${\bf y}_{\rm TF}^n$ and ${\bf x}_{\rm TF}^n$ are the $n$-th received and transmitted OFDM symbol, respectively, ${\bf w} \in {\cal CN}(0,\sigma^2_{\bf w} {\bf I}_N)$ refers to the additional white Gaussian noise (AWGN), and ${\bf F} \in {\mathbb C}^{M \times M}$ is the DFT matrix. {Besides, $k_i = NT_{\rm sym} \nu_i$ and $l_i = M\Delta f \tau_i$ are the normalized Doppler and delay w.r.t. the TF domain resources, respectively.} Considering the impact of CP, we further denote $\kappa_i = NT \nu_i$, with $\kappa_i = \frac{T}{T_{\rm sym}}k_i $. Besides, ${\bf \Pi}^{l_i}$ in (\ref{OFDMIOTD}) refers to the cyclic shift matrix corresponding to the delay $\tau_i$, given as:
\begin{equation}
{\bf \Pi}^{l_i}= ({\bf 0}_{1\times (M-1)} 1;{\bf I}_{(M-1)\times (M-1)} {\bf 0}_{(M-1)\times1})^{l_i}.
\end{equation}
Additionally, ${\bf{\Lambda }}_n^{\left( {{k_i}} \right)}$ refers to the phase shift matrix corresponding to the Doppler shift $\nu_i$:
\begin{equation}
{\bf{\Lambda }}_n^{\left( {{k_i}} \right)} =  {e^{j2\pi \frac{k_in}{N}}}{\rm{diag}}( {{e^{\frac{{j2\pi {\kappa_i}(0)}}{{NM}}}},...,{e^{\frac{{j2\pi {\kappa_i}(M - 1)}}{{NM}}}}}).
\end{equation}

In (\ref{OFDMIOTD}), $\sum^{P}_{i = 1} h_i{\bf F}{\bf \Pi}^{l_i} {\bf \Lambda}_{n}^{(k_i)}{\bf F}^{H}$ represents the TF domain equivalent matrix of the $n$-th OFDM symbol, which is denoted as ${\bf H}_{\rm TF}^n \in {\mathbb C}^{M \times M}$. Under the impact of Doppler shifts of multipath, ${\bf{\Lambda }}_n^{\left( {{k_i}} \right)}$ is not an identity matrix and ${\bf H}_{\rm TF}^n$ is not diagonal, leading to the ICI.\par

Furthermore, using the property that the cyclic matrix can be diagonalized by unity matrixes, it can be proved that:
\begin{equation}
\begin{aligned}
{\bf{H}}_{{\rm{TF}}}^n = \sum\limits_{i = 1}^P {{h_i}} {e^{j2\pi \frac{nk_i}{N}}}{\rm diag}\left( {{e^{\frac{{ - j2\pi (0){l_i}}}{M}}}},...,{{e^{\frac{{ - j2\pi (M-1){l_i}}}{M}}}} \right){\bf A}_i,
\end{aligned}
\label{CSF_equ}
\end{equation}
where ${\bf A}_i$ refers to the ICI matrix {\color{black}incurred} by the $i$-th path, whose elements are given by:
\begin{equation}
{{\bf A}_i[p,q]} = \frac{{\sin \left( {\pi \left( {q - p + \frac{{{\kappa_i}}}{N}} \right)} \right)}}{{M\sin \left( {\frac{\pi }{M}\left( {q - p + \frac{{{\kappa_i}}}{N}} \right)} \right)}} \cdot {e^{j\pi \left( {q - p + \frac{{{\kappa_i}}}{N}} \right)\left( {1 - \frac{1}{M}} \right)}},
\label{A_11}
\end{equation}
where $p,q\in\{0,1,...,M-1\}$. Based on $\sum_{m=0}^{M-1} \frac{1}{\sin ^2\left(\frac{\pi}{M}(m+x)\right)}=\frac{M^2}{\sin ^2(\pi x)}$, we can find that $\sum^{M-1}_{p=q+m,m=0} |{{\bf A}_i[p,q]}|^2 = 1$ and $|{{\bf A}_i[0,0]}|^2$ is maximum among $|{{\bf A}_i[p,q]}|^2$.
\section{CSF Characterization Based on Discrete TF Domain Pilots}
\label{section III}
In this section, we will detail the characteristic of the CSF estimated based on the discrete pilots in DD-a-OFDM.
\subsection{Pilot Arrangements in OFDM Systems}
\label{pilotpattern}



\begin{figure}[!h]
  \centering
  \includegraphics[width=.48\textwidth]{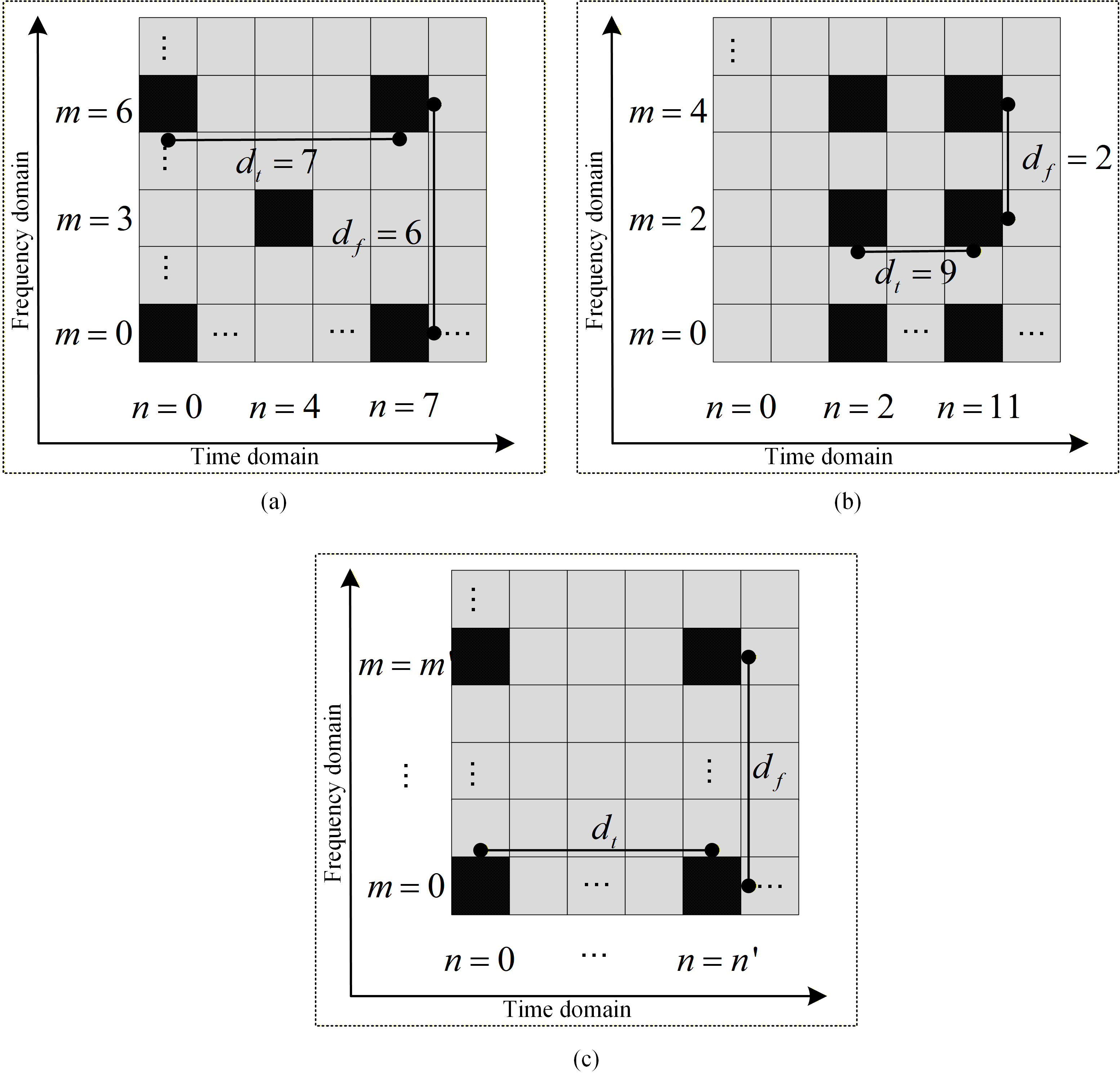}
  \caption{Pilot patterns in OFDM systems: (a) an example of CRS pattern in LTE, (b) an example of DMRS pattern in 5G-NR for high-speed mobility support, and (c) the general pilot pattern considered by DD-a-OFDM. The black and grey bins {\color{black}refer} to the pilot and data symbols, respectively.}\label{FIG2}
  \vspace{-1em}
\end{figure}

In commercial OFDM systems, pilot symbols are discretely distributed among $NM$ TF domain resource elements (REs) according to channel coherence properties \cite{DTse}. As illustrated in Fig. \ref{FIG2}(a) and Fig. \ref{FIG2}(b), LTE systems employ cell reference signals (CRS) while 5G NR systems utilize demodulation reference signals (DMRS) as their respective pilot patterns. Without loss of generality, under the OFDM framework, we denote the pilot spacing in time and frequency domains as $d_t$ and $d_f$ respectively, satisfying ${\rm mod}(N,d_t) = 0$ and ${\rm mod}(M,d_f) = 0$, as depicted in Fig. \ref{FIG2}(c). All pilots maintain equal energy $E_{\rm p}$ with data symbols.\par

\subsection{CSF Characterization}
By separating the desired transmitted signals and the ICI terms of (\ref{CSF_equ}) in (\ref{OFDMIOTD}), the TF domain I/O relationship can be rewritten as:
\begin{equation}
\begin{aligned}
&{{\bf{y}}_{{\rm{TF}}}}[m,n] {=} \sum\limits_{i {=} 1}^P{h_i}{e^{j2\pi {\nu _i}nT_{\rm sym}}}\!{e^{{{ {-} j2\pi \tau_i m\Delta f}}}}\!{{\bf A}_i[m,m]}\!{{\bf{x}}_{{\rm{TF}}}}[m,n] \\
& {+}\underbrace {\sum\limits_{m' \ne m} {{h_i}{e^{j2\pi {\nu _i}nT_{\rm sym}}}{e^{ - j2\pi {\tau _i}m\Delta f}}} {{\bf A}_i[m,m']}{{\bf{x}}_{{\rm{TF}}}}[m',n]}_{{{\bf{I}}_{i,{\rm TF}}}[m,n]}\\
& {+} {\bf{w}}[m,n],
\end{aligned}
\label{TF_IO}
\end{equation}
where, ${{{\bf{I}}_{i,{\rm TF}}}[m,n]}$ refers to the ICI suffered by the symbol located on the $n$-th OFDM symbol and the $m$-th subcarrier. To obtain CSF, we can estimate CTF firstly, and then transform the estimated CTF into DD domain via the symplectic finite Fourier transform (SFFT). The detailed steps {\color{black}involve}:\par
{\bf Step 1}: First, based on the discrete pilots defined in Section \ref{pilotpattern}, the CTF of pilot symbols in (\ref{TF_IO}) can be estimated. If the least square (LS) principle is adopted, we can obtain:
\begin{equation}
\begin{aligned}
  &\hat h_{{\rm{TF}}}^{{\rm{Pilot}}}[m,n] = {{\bf{y}}_{{\rm{TF}}}}[m,n]/{{\bf{x}}_{{\rm{TF}}}}[m,n]\\
  &{=}\! \sum\limits_{i \! = \! 1}^P \!\!{{h_i} {e^{j2\pi \!{\nu _i}nT_{\rm sym}}}\! {e^{\! - \!j2\pi {\tau _i}m\Delta f}}\!\! {\bf A}_i[m,m]}\! {+} \frac{{{{\bf{I}}_{i,{\rm TF}}}[m,n]}}{{{{\bf{x}}_{{\rm{TF}}}}\! [m,n]}}\! {+} {\tilde {\bf w}}\! [m,n],
\end{aligned}
\label{TF_LS}
\end{equation}
where $m\in\{0,d_f-1,...,{M}-1\}$, $n \in \{0,d_t-1,...,N-1\}$. Besides, ${\tilde {\bf w}} [m,n]$ is the equivalent AWGN with ${\tilde {\bf w}} [m,n] \sim {\cal CN}(0,\sigma^2_{\bf w}{\mathbb E}\{\frac{1}{{|{{\bf{x}}_{{\rm{TF}}}}[m,n]|^2}}\})$. For brevity, ${\mathbb E}\{\frac{1}{{|{{\bf{x}}_{{\rm{TF}}}}[m,n]|^2}}\}$ is denoted by $\sigma^2_{1/x}$, where $\sigma^2_{1/x} = 1/E_{\rm p}$ for 4-QAM modulation. Then, the estimated discrete CTF of $NM$ TF domain REs is denoted by:
\begin{equation}
\begin{aligned}
&{\hat h}^{\rm Discrete}_{\rm TF}[m,n] \!\!=\! \begin{cases} \!{\hat h}^{\rm Pilot}_{\rm TF}[m'd_f,n'd_t],\!\!& \begin{aligned}& m'\!\!\in\!\{0,\!1,...,\!\frac{M}{d_f}\!-\!1\},\\
& n'\!\!\in\!\{0,\!1,...,\!\frac{N}{d_t}\!-\!1\}\end{aligned} \\
0, & {\rm elsewhere},
\end{cases}
\end{aligned}
\label{CTF_Dis}
\end{equation}

{\bf Step 2}: Afterwards, the periodic observation of the original CSF can be obtained by applying SFFT on $\hat h_{{\rm{TF}}}^{{\rm{Discrete}}}[m,n]$ as:
\begin{equation}
\begin{aligned}
& \hat h_{{\text{DD}}}^{{\text{Periodic}}}\left[ {k,l} \right] = DF{T_N}\{ IDF{T_M}\{ \hat h_{{\rm{TF}}}^{{\rm{Discrete}}}[m,n]\} \}\\
& = \sum_{i=1}^P h_i \underbrace{\sum_{m^{\prime}=0}^{\frac{M}{d_f}-1} \frac{d_f}{\sqrt{{M}}} e^{-j 2 \pi m^{\prime} d_f \frac{\left(l_i-l\right)}{M}}}_{\mathcal{R}_{\text {delay }}^{\text {Periodic }}\left(l_i, l\right)} \underbrace{\sum_{n^{\prime}=0}^{\frac{N}{d_t}-1} \frac{d_t}{\sqrt{{N}}} e^{j 2 \pi n^{\prime} d_t \frac{\left(k_i-k\right)}{N}}}_{\mathcal{R}_{\text {Doppler }}^{\text {Periodic }}\left(k_i, k\right)},\\
& \times {{\bf A}_i[0,0]} +\sum\limits_{i = 1}^P  {{\bf I}_{i,{\rm DD}}}[k,l] + {\tilde {\bf w}}[k,l],\\
\label{CSF_pe}
\end{aligned}
\end{equation}
where ${\bf I}_{i,{\rm DD}} \in {\mathbb C}^{N \times M}$ in (\ref{CSF_pe}) refers to the equivalent ICI in DD domain. Besides, ${{\cal R}^{\rm Periodic}_{\rm delay}(l_i,l)}$ and ${{\cal R}^{\rm Periodic}_{\rm Doppler}(k_i,k)}$ refers to the sampling term associated with the delay $\tau_i$ and Doppler $\nu_i$ for the desired signal component, respectively. Under the assumption that $l_i$ is on-grid w.r.t. $M\Delta f$ and $k_i$ is off-grid w.r.t. $NT_{\rm sym}$, ${{\cal R}^{\rm Periodic}_{\rm delay}(l_i,l)}$ and ${{\cal R}^{\rm Periodic}_{\rm Doppler}(k_i,k)}$ are given by:
\begin{equation}
\begin{aligned}
{{\cal R}^{\rm Periodic}_{\rm delay}(l_i,l)}  &= \sum\limits_{m' \!=\! 0}^{\frac{M}{d_f} \!-\! 1} \! \frac{{{d_f}}}{{\sqrt M }}{{e^{ \!-j2\pi m'd_f\frac{{\left( {{l_i}\! -\! l} \right)}}{M}}}}\\
& =\!\! \begin{cases}
\sqrt{M}, & \text{if } {\rm mod}(l_i {-} l,\frac{M}{d_f})\!=\!0, \\
0, & \text{otherwise}.
\end{cases}
\end{aligned}
\label{R_d}
\end{equation}
\begin{equation}
\begin{aligned}
 &{{\cal R}^{\rm Periodic}_{\rm Doppler}(k_i,k)} = {\!\sum\limits_{n' \!=\! 0}^{\frac{N}{{{d_t}}}\!-\!1}\! \frac{{{d_t}}}{{\sqrt N }}{{e^{ j2\pi n'd_{t}\frac{{\left( {{k_i} \!-\! k} \right)}}{N}}}} }\\
& =\!\! \begin{cases}
\sqrt{N}, & \!\!\text{if } {\rm mod}(k_i {-} k,\frac{N}{d_t})\!=\!0, \\
\!\!\frac{{{d_t}}}{{\sqrt N }}{e^{ j\pi \left( {{k_i} {-} k} \right)(1 {-} \frac{{{d_t}}}{N})}}\!\frac{{\sin \left( {\pi \left( {{k_i} {-} k} \right)} \right)}}{{\sin ( {\pi {d_t}\frac{{\left( {{k_i} {-} k} \right)}}{N}} )}}
, & \!\!\text{otherwise}.
\end{cases}
\end{aligned}
\label{R_D}
\end{equation}
Therein, the normalization factors in ${{\cal R}^{\rm Periodic}_{\rm delay}(l_i,l)}$ and ${{\cal R}^{\rm Periodic}_{\rm Doppler}(k_i,k)}$ are $\frac{d_f}{\sqrt{M}}$ and $\frac{d_t}{\sqrt{N}}$, respectively, to make the peak values of $\hat h_{{\text{DD}}}^{{\text{Periodic}}}\left[ {k,l} \right]$ similar to those of the real CSF.

{\color{black}
\begin{remark}
{\color{black}The unambiguous delay and Doppler ranges supported by the periodic CSF observation are determined by the pilot intervals $(d_f, d_t)$. To avoid aliasing, the delay and Doppler shift must satisfy $\tau_{\max} \le 1/(d_f \Delta f)$ and $\nu_{\max} \le 1/(2 d_t T_{\rm sym})$. For a typical 5G NR configuration with $\Delta f = 15$ kHz, $d_f = 4$, and $d_t = 4$, the system supports a maximum delay of $\tau_{\max} \approx 16.67$ $\mu$s and a maximum Doppler shift of $\nu_{\max} \approx 1.875$ kHz. These supportable limits translate to a relative speed of over $330$ km/h at $f_c = 6$ GHz and approximately $960$ km/h at $f_c = 2.1$ GHz, enveloping the maximum delay and Doppler spreads of realistic scattering environments for high-mobility railway and vehicular communication systems, such as the 3GPP TDL-C channel model.}\par
\label{remark1}
\end{remark}
}

\subsection{Necessity of Precise DD domain Parameter Estimations}
\label{DDPE}
{\color{black}Similar to the relationship between the classical time and frequency domain signals, the CSF can be regarded as the ``frequency response'' of the sampled CTF. Under the Nyquist-compliant pilot spacing discussed in Remark \ref{remark1}, one period of the periodic CSF observation in (\ref{CSF_pe}) spans $k\in [-\frac{N}{2d_t},\frac{N}{2d_t}-1]$ and $l \in [0,\frac{M}{d_f}-1]$. Theoretically, as adopted in 4G and 5G systems, one might apply TF domain interpolation (e.g., linear or ideal 2D low-pass filtering) to estimate the CTF at data symbols, which is equivalent to obtaining the CSF by filtering the periodic observation (\ref{CSF_pe}).

However, such traditional approaches are insufficient for precise CSF estimation in high-mobility scenarios. Although the pilot intervals satisfy the Nyquist criteria, the off-grid Doppler and delay cause the sampled CSF energy to spread across the DD grid according to the Dirichlet-kernel in (\ref{R_D}). This energy leakage inevitably spills over the boundaries of the observation span defined in Remark \ref{remark1}, creating aliasing that simple LPF-based filtering or TF domain interpolation cannot eliminate. Consequently, these methods fail to resolve the fractional parameters required for accurate ICI-aware equalization.}

\begin{figure}
  \centering
  \includegraphics[width=.48\textwidth]{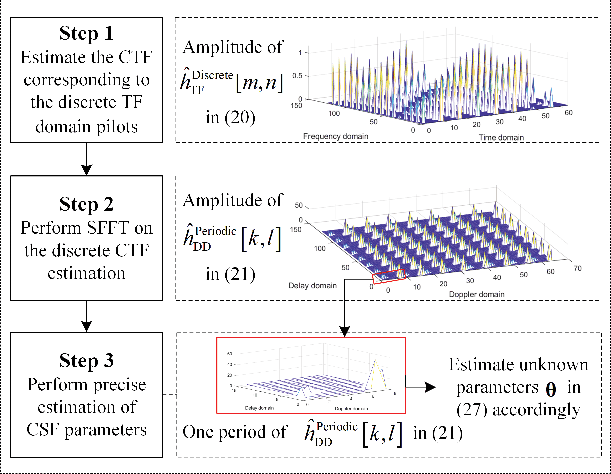}
  \caption{CSF estimation procedure based on discrete pilots in DD-a-OFDM.}\label{FIG3}
  \vspace{-1em}
\end{figure}

Therefore, we propose a DD domain signal processing framework to achieve precise CSF estimation. The procedure is summarized as: (1) Estimate the CTF at discrete TF pilot positions; (2) Perform SFFT on the discrete CTF estimations; (3) Perform precise estimation of CSF parameters based on the obtained observation. The overall procedure is visualized in Fig. \ref{FIG3}. We note that these three steps are general and do not specify mathematical methods. The first 2 steps have been detailed in (\ref{TF_LS})-(\ref{R_D}), and the last step will be discussed in Section IV.\par

\subsection{Impact of ICI on CSF estimation in DD-a-OFDM}
\label{ICI}
Before estimating the precise DD domain channel parameters according to (\ref{CSF_pe}), we note that the distribution of ${\bf I}_{i,\rm{DD}}[k,l]$ warrants careful analysis, which has been neglected in \cite{wlc} but will limit the CSF estimation accuracy in the high mobility conditions. In detail, ${\bf I}_{i,\rm{DD}}[k,l]$ can be expanded as: 
\begin{equation}
\begin{aligned}
&{\bf I}_{i,{\rm DD}}[k,l] \\
& {=} \sum\limits_{m' {=} 0}^{\frac{M}{{{d_f}}} {-} 1} \!\! \frac{{{d_f}}}{{\sqrt M }}\!{e^{ j2\pi m'{d_f}\!\frac{{{l}}}{M}}}\!\!\sum\limits_{n' = 0}^{\frac{N}{{{d_t}}} - 1}\!\!\!\frac{{{d_t}}}{{\sqrt N }}{e^{-j2\pi n'{d_t}\frac{{{k}}}{N}}}\frac{{{\bf{I}}_{i,{\rm TF}}}[m'd_f,n'd_t]}{{{{\bf{x}}_{{\rm{TF}}}}\! [m'd_f,n'd_t]}}\\
& {\mathop  = \limits^{(a)} } {{h_i}}\sum\limits_{m' {=} 0}^{\frac{M}{{{d_f}}} {-} 1} \!\! \frac{{{d_f}}}{{\sqrt M }} \!{e^{ j2\pi m'{d_f}\!\frac{{{l}}}{M}}}{{e^{ - j2\pi {\tau _i}m'{d_f}\Delta f}}}\!\!\!\!\!\sum\limits_{m'' {\ne} m'{d_f}}\!\!\!\!\! {\bf A}_i[m'{d_f},m'']\\
& \times \underbrace{\sum\limits_{n' = 0}^{\frac{N}{{{d_t}}} - 1}  \frac{{{d_t}}}{{\sqrt N }}{e^{-j2\pi n'{d_t}\frac{{{k}}}{N}}}{e^{j2\pi {\nu _i}n'{d_t}T_{\rm sym}}}\frac{{{{\bf{x}}_{{\rm{TF}}}}[m'',n'{d_t}]}}{{{{\bf{x}}_{{\rm{TF}}}}[m'{d_f},n'{d_t}]}}}_{\omega^{m'',m'}_{\rm FD}}.\\
\label{I_DD}
\end{aligned}
\end{equation}
Therein, equation ``(a)'' is obtained by substituting ${{\bf{I}}_{i,{\rm TF}}}[m,n]$ of (\ref{TF_IO}) inside. According to equation (a) of (\ref{I_DD}), we note that ${\bf I}_{i,\rm{DD}}[k,l]$ physically represents the IDFT of the ICI impacted frequency-Doppler domain signal, where the frequency-Doppler domain signal is denoted by ${\omega^{m'',m'}_{\rm FD}}$.\par

Denote the ratio of data and pilot symbols $\frac{{{{\bf{x}}_{{\rm{TF}}}}[m'',n'{d_t}]}}{{{{\bf{x}}_{{\rm{TF}}}}[m'{d_f},n'{d_t}]}}$ as a random variable ${\mathcal X}$, we can deduce that the term $\omega^{m'',m'}_{\rm FD}$ will follow a complex Gaussian distribution when $N/d_t$ is sufficiently large and ${\mathcal X}$ is independently and identically distributed (i.i.d.), according to the central limit theorem (CLT) \cite{bxxx}. By assuming that the data symbols ${{{{\bf{x}}_{{\rm{TF}}}}[m'',n'{d_t}]}}$ are selected randomly selected from the QAM modulation alphabet, ${{{{\bf{x}}_{{\rm{TF}}}}[m'',n'{d_t}]}}$ can be regarded as i.i.d. variables. Meanwhile, let the pilots ${{{{\bf{x}}_{{\rm{TF}}}}[m'{d_f},n'{d_t}]}}$ {\color{black}be} randomly selected from the same alphabet as data symbols, the i.i.d. assumption of ${\mathcal X}$ will be valid. By doing so, we can obtain that ${\mathbb E}\{{{{{\bf{x}}_{{\rm{TF}}}}[m'',n'{d_t}]}}\} = 0$ holds for both data and pilot symbols. Moreover, the mean and variance of i.i.d. variable $\cal X$ are given by ${\mathbb E}\{{\cal X}\} = 0$ and ${\rm Var}\{{\cal X}\} = {E_{\rm p}}{{\mathbb E}\{\frac{1}{{|{{\bf{x}}_{{\rm{TF}}}}[m'{d_f},n'{d_t}]|^2}}\}} =E_{\rm p} \sigma^2_{1/x}$, respectively. Based on the above, the distribution of the frequency-Doppler domain signal $\omega^{m'',m'}_{\rm FD}$ can be written as $\omega^{m'',m'}_{\rm FD} \sim \mathcal{CN}(0,d_t E_{\rm p}\sigma^2_{1/x})$. For instance, with 4-QAM modulation where $\sigma^2_{1/x} = 1/E_{\rm p}$, the above distribution simplifies to $\omega^{m'',m'}_{\rm FD} \sim \mathcal{CN}(0,d_t)$.\par

{\color{black}Moreover, since ${\bf I}_{i,\rm{DD}}[k,l]$ represents the sum of i.i.d. Gaussian $\omega_{\rm FD}^{m'',m'}$, ${\bf I}_{i,\rm{DD}}[k,l]$ can also be modeled as a complex Gaussian variable}. {\color{black} By utilizing the unitary energy property of the leakage matrix derived in (\ref{A_11}), i.e., $\sum_{m=0}^{M-1} |{\bf A}_i[q+m, q]|^2 = 1$, the total leakage energy excluding the self-interference is invariant of the frequency index $m'$, yielding $\sum_{m'' \ne m'{d_f}} |{\bf A}_i[m'{d_f},m'']|^2 = 1 - |{\bf A}_i[0,0]|^2$. This ensures that ${\bf I}_{i,\rm{DD}}[k,l]$ follows an identical complex Gaussian distribution across the DD domain grid. Furthermore, the strict independence of ${\bf I}_{i,\rm{DD}}[k,l]$ is partially compromised due to the term $\sum\limits_{m'' {\ne} m'{d_f}}\!\!\!\!\! {\bf A}_i[m'{d_f},m'']{\omega^{m'',m'}_{\rm FD}}$ in (\ref{I_DD}). However, according to (\ref{A_11}), ${\bf A}_i[m'{d_f},m'']$ exhibit a Dirichlet-kernel (Sinc-like) shape, due to the rapid decay of the Sinc side-lobes, this cross-bin leakage is localized. Such weak spatial correlation allows the interference to be treated as approximately independent without loss of analytical rigor.} Hence, we can model ${\bf I}_{i,{\rm DD}}[k,l]$ in (\ref{I_DD}) as {\color{black} i.i.d. Gaussian variables, i.e., ${\bf I}_{i,{\rm DD}}[k,l] \sim {\cal CN}(0,|h_i|^2 (1-|{\bf A}_i[0,0]|^2) d_t d_f E_{\rm p}\sigma^2_{1/x})$}. This leads to the compact representation of (\ref{CSF_pe}):
\begin{equation}
\begin{aligned}
\hat h_{{\text{DD}}}^{{\text{Periodic}}}\left[ {k,l} \right] &= \sum\limits_{i = 1}^P {{\tilde h_i}} {\cal R}_{{\text{delay}}}^{{\text{Periodic}}}({l_i},l){\cal R}_{{\text{Doppler}}}^{{\text{Periodic}}}({k_i},k)\\
&+ {\bf v}[k,l].
\end{aligned}
\label{h_DD}
\end{equation}
Therein, $\tilde h_i = h_i{{\bf A}_i[0,0]}$. Besides, the ${\bf v}[k,l]$ refers to the equivalent noise composed of DD domain ICI and AWGN, with ${\bf v}[k,l] \sim {\cal CN}(0,\sigma^2_{\bf v})$ and:
\begin{equation}
\sigma^2_{\bf v} = \sum^{P}_{i=1}|h_i|^2 (1-|{\bf A}_i[0,0]|^2)E_{\rm p}\sigma^2_{1/x}d_t d_f+{{{\sigma^2_{\bf w}}}}{{{\sigma^2_{1/x}}}}d_td_f,
\label{sigma_v}
\end{equation}
which is obtained by {\color{black} aggregating the AWGN variance and the derived ICI variance.}

\begin{remark}
The proposed DD-a-OFDM system design reveals two crucial insights regarding CSF estimation. First, the CSF estimation accuracy fundamentally depends on pilot spacing parameters $d_t$ and $d_f$ through the equivalent noise variance $\sigma^2_{\bf v}$ in (\ref{sigma_v}), distinguishing DD-a-OFDM from conventional pilot-based OFDM systems. This dependence manifests in the trade-off between pilot overhead and estimation precision. Second, while inherent ICI effects persist due to the OFDM waveform's interaction with LTV channels, our analysis demonstrates their transformation into Gaussian-distributed interference in the DD domain when $M/d_f$ and $N/d_t$ are relatively large, as analysed before (\ref{h_DD}). This property provides key advantages: (1) elimination of non-Gaussian ICI effects that plague traditional CTF estimation in OFDM systems, and (2) simplified mitigation of the resulting Gaussian-distributed interference within the DD-a-OFDM framework.
\end{remark}

\subsection{CRLB Derivation}
\label{SECTIONXXX}
\begin{figure*}
\setcounter{equation}{34}
\begin{equation}
\begin{aligned}
&\mathbf{I}(\boldsymbol{\theta}, \boldsymbol{x}) = \frac{2}{\sigma_{\bf v}^2}\left({\begin{array}{*{20}{c}}
{NM}&0&0&0\\
0&{{{| {{{\tilde h}_i}} |}^2}NM}&{\pi {{| {{{\tilde h}_i}} |}^2}M( {N{-}{d_t}} )}&{{-}\pi {{| {{{\tilde h}_i}} |}^2}N( {M{-}{d_f}} )}\\
0&{\pi {{| {{{\tilde h}_i}} |}^2}M( {N{-}{d_t}} )}&{{\pi ^2}{{| {{{\tilde h}_i}} |}^2}M( {N{-}{d_t}} )( {\frac{{4N{-}2{d_t}}}{{3N}}} )}&{{-}{\pi ^2}{{| {{{\tilde h}_i}} |}^2}( {N{-}{d_t}} )( {M{-}{d_f}} )}\\
0&{{-}\pi {{| {{{\tilde h}_i}} |}^2}N( {M{-}{d_f}} )}&{{-}{\pi ^2}{{| {{{\tilde h}_i}} |}^2}( {N{-}{d_t}} )( {M{-} {d_f}} )}&{{\pi ^2}{{| {{{\tilde h}_i}} |}^2}N( {M{-}{d_f}} )( {\frac{{4M{-}2{d_f}}}{{3M}}} )}
\end{array}} \right).
\end{aligned}
\label{I_og}
\end{equation}
\setcounter{equation}{26}
\hrule
\end{figure*}
Before estimating precise DD domain channel parameters based on (\ref{h_DD}), since (\ref{h_DD}) has not {\color{black}been} shown in the conventional OFDM and OTFS systems and is unique for the DD-a-OFDM system, the theoretical bounds for CSF estimation are required. Therefore, we will derive the CRLB of $\tilde h_i$, $l_i$, $k_i$ estimations for (\ref{h_DD}) subsequently. {\color{black}To beginning with}, the unknown parameters are denoted as $\boldsymbol{\theta} = \{\left|{\tilde h}_i\right|,\phi_i,k_i,l_i\}_{4P\times 1}$, where $\phi_i = \angle {\tilde h}_i$. Accordingly, the log-likelihood function for CSF estimation based on (\ref{h_DD}) is given by:
\begin{equation}
\begin{aligned}
& {\mathcal L}\left( {\hat h_{{\text{DD}}}^{{\text{Periodic}}}|\boldsymbol{\theta}} \right) = \prod\limits_{l = 0}^{\frac{M}{{{d_f}}} - 1} {\prod\limits_{k = 0}^{\frac{N}{{{d_t}}} - 1} {\frac{1}{{\sqrt {2\pi \sigma _{\bf v}^2} }}} } \\
& \times {e^{ - \frac{1}{{2\sigma _{\bf v}^2}}{{\left| {\hat h_{{\text{DD}}}^{{\text{Periodic}}}\left[ {k,l} \right] - \sum\limits_{i = 1}^P {|{{\tilde h}_i}|e^{j\angle {{\tilde h}_i}}} {{\cal R}^{\rm Periodic}_{\rm delay}(l_i,l)}{\cal R}^{\rm Periodic}_{\rm Doppler}(k_i,k) } \right|}^2}}}.
\end{aligned}
\label{llf}
\end{equation}

For notational simplicity, we define {\color{black} the noise-free signal component of the $i$-th path as}:
\begin{equation}
s_i[k, l]=\left|{\tilde h}_i\right| e^{j \phi_i} {{\cal R}^{\rm Periodic}_{\rm delay}(l_i,l)}{\cal R}^{\rm Periodic}_{\rm Doppler}(k_i,k).
\label{equ_s}
\end{equation}
{\color{black} Accordingly, the total noise-free observation at the $(k,l)$-th DD domain grid is denoted as $s[k, l] = \sum_{i=1}^P s_i[k, l]$. Let $\theta_{i_p}$ denote the $p$-th parameter ($p \in \{1, 2, 3, 4\}$) of the $i$-th path in $\boldsymbol{\theta}$, with its global index mapped as $i_p = 4(i-1) + p$. Similarly, $\theta_{j_q}$ corresponds to the $j$-th path with global index $j_q = 4(j-1) + q$. According to the log-likelihood function in (\ref{llf}), the fully coupled} Fisher information matrix (FIM) $\mathbf{I}(\boldsymbol{\theta}, \boldsymbol{x}) {\color{black} \in \mathbb{R}^{4P \times 4P}}$ is defined by evaluating the inner products of the signal derivatives \cite{Gaudio}:
\begin{equation}
{\color{black} \mathbf{I}(\boldsymbol{\theta}, \boldsymbol{x})[i_p, j_q] = \frac{2}{\sigma_{\bf v}^2} \operatorname{Re}\left\{\sum\limits^{\frac{M}{d_f}-1}_{l=0}\sum\limits^{\frac{N}{d_t}-1}_{k=0}\left[\frac{\partial s[k, l]}{\partial \theta_{i_p}}\right]^* \frac{\partial s[k, l]}{\partial \theta_{j_q}}\right\}.}
\label{Fisher_I_full}
\end{equation}
{\color{black} Considering that the unknown parameter $\theta_{i,p}$ solely affects the signal component of the $i$-th path $s_i[k,l]$, i.e., $\frac{\partial s[k,l]}{\partial \theta_{i_p}} = \frac{\partial s_i[k,l]}{\partial \theta_{i_p}}$, \eqref{Fisher_I_full} can be simplified as:}
\begin{equation}
{\color{black} \mathbf{I}(\boldsymbol{\theta}, \boldsymbol{x})[i_p, j_q] = \frac{2}{\sigma_{\bf v}^2} \operatorname{Re}\left\{\sum\limits^{\frac{M}{d_f}-1}_{l=0}\sum\limits^{\frac{N}{d_t}-1}_{k=0}\left[\frac{\partial s_i[k, l]}{\partial \theta_{i_p}}\right]^* \frac{\partial s_j[k, l]}{\partial \theta_{j_q}}\right\}.}
\end{equation}

{\color{black} When $i \neq j$, the off-diagonal elements capture the inter-path interference (IPI) coupling caused by the non-orthogonality of the fractional basis components. However, due to the rapid decay of the Dirichlet-kernel side-lobes in \eqref{R_d} and \eqref{R_D}, the cross-correlation energy between different paths is small. By adopting a well-justified block-diagonal approximation, the cross-path elements are approximately zero, yielding the decoupled FIM:}
{\color{black}\begin{equation}
\begin{aligned}
&\mathbf{I}(\boldsymbol{\theta}, \boldsymbol{x})[i_p, j_q] \\
&{\approx \begin{cases}
\frac{2}{\sigma_{\bf v}^2} \operatorname{Re}\left\{\sum\limits^{\frac{M}{d_f}-1}_{l=0}\sum\limits^{\frac{N}{d_t}-1}_{k=0}\left[\frac{\partial s_i[k, l]}{\partial \theta_{i_p}}\right]^* \frac{\partial s_i[k, l]}{\partial \theta_{i_q}}\right\}, & \text{if } i = j, \\
0, & \text{if } i \neq j.
\end{cases}}
\end{aligned}
\label{Fisher_I}
\end{equation}}

{\color{black} Therefore, the CRLB for $\boldsymbol{\theta}(i_p)$ can be obtained as:}
\begin{equation}
{\rm CRLB}(\boldsymbol{\theta}(i_p)) = {\rm diag}(\mathbf{I}(\boldsymbol{\theta}, \boldsymbol{x})^{-1})[i_p,i_p].
\label{crlb111}
\end{equation}

In detail, the basic derivation functions used for CRLB derivations are given by:
\begin{subequations}
\begin{equation}
\begin{aligned}
&\frac{\partial s_i[k, l]}{\partial \left|{\tilde h}_i\right|} = e^{j \phi_i} {{\cal R}^{\rm Periodic}_{\rm delay}(l_i,l)}{\cal R}^{\rm Periodic}_{\rm Doppler}(k_i,k),
\end{aligned}
\label{s_1}
\end{equation}
\begin{equation}
\begin{aligned}
&\frac{\partial s_i[k, l]}{\partial \phi_i} = j \left|{\tilde h}_i\right|e^{j \phi_i} {{\cal R}^{\rm Periodic}_{\rm delay}(l_i,l)}{\cal R}^{\rm Periodic}_{\rm Doppler}(k_p,k),
\end{aligned}
\label{s_4_2}
\end{equation}
\begin{equation}
\begin{aligned}
\frac{\partial s_i[k, l]}{\partial l_i} &= \left|{\tilde h}_i\right| e^{j \phi_i} \sum\limits_{m' = 0}^{\frac{M}{{{d_f}}} - 1}  { - j2\pi m'{d_f}\frac{1}{M}} \frac{{{d_f}}}{{\sqrt M }}{e^{ - j2\pi m'{d_f}\frac{{\left( {{l_i} - l} \right)}}{M}}}\\
& \times {\cal R}^{\rm Periodic}_{\rm Doppler}(k_i,k),
\end{aligned}
\end{equation}
\begin{equation}
\begin{aligned}
\frac{\partial s_i[k, l]}{\partial k_i} &= \left|{\tilde h}_i\right| e^{j \phi_i} {\!\sum\limits_{n' \!=\! 0}^{\frac{N}{{{d_t}}}\!-\!1}\! j2\pi n'd_{t}\frac{1}{N}\frac{{{d_t}}}{{\sqrt N }}{{e^{ j2\pi n'd_{t}\frac{{\left( {{k_i} \!-\! k} \right)}}{N}}}} }\\
& \times {\cal R}^{\rm Periodic}_{\rm delay}(l_i,l).
\end{aligned}
\label{s_4}
\end{equation}
\label{derivations}
\end{subequations}

Crucially, the FIM in (\ref{Fisher_I}) is independent of the off-grid channel conditions, i.e., fractional $k_i$, by virtue of Parseval's theorem \cite{bxxx}. Taking the phase-related element $I(\boldsymbol{\theta}, \boldsymbol{x})[2, 2]$ with $i=1$ as an example, substituting \eqref{s_4_2} into \eqref{Fisher_I} yields:
\begin{equation}
\begin{aligned}
&I(\boldsymbol{\theta}, \boldsymbol{x})[2, 2]\\
 &= \frac{2}{\sigma_{\bf v}^2} \left|{\tilde h}_i\right|^2 \sum\limits^{\frac{M}{d_f}-1}_{l=0} |{{\cal R}^{\rm Periodic}_{\rm delay}(l_i,l)}|^2 \sum\limits^{\frac{N}{d_t}-1}_{k=0} |{\cal R}^{\rm Periodic}_{\rm Doppler}(k_i,k)|^2 \\
&\stackrel{\text{Parseval}}{=} \frac{2}{\sigma_{\bf v}^2} \left|{\tilde h}_i\right|^2 M \left[ \sum_{n' = 0}^{\frac{N}{d_t}-1} \left| \frac{d_t}{\sqrt N} e^{ j2\pi n'd_{t}\frac{k_i}{N}} \right|^2 \right]\\
& = \frac{2}{\sigma_{\bf v}^2} \left|{\tilde h}_i\right|^2 NM,
\end{aligned}
\label{I_22}
\end{equation}
therein, the equality ``$\stackrel{\text{Parseval}}{=}$'' by substituting \eqref{R_d} and applying Parseval's theorem \cite{bxxx}. Specifically, by treating ${\cal R}^{\rm Periodic}_{\rm Doppler}(k_i,k)$ in the Doppler domain as the IDFT of ${\frac{{{d_t}}}{{\sqrt N }}{{e^{ j2\pi n'd_{t}\frac{{{k_i}}}{N}}}}}$ in the time domain according to (\ref{R_D}),  Parseval's theorem ensures that the energy of ${\cal R}^{\rm Periodic}_{\rm Doppler}(k_i,k)$ is equal to the energy of its Fourier transform, i.e., ${\frac{{{d_t}}}{{\sqrt N }}{{e^{ j2\pi n'd_{t}\frac{{{k_i}}}{N}}}}}$, owing to the unitary property of the Fourier transform. Interestingly, \eqref{I_22} can also be obtained by assuming the Doppler shift is on-grid, similar to \eqref{R_d}, which simplifies the derivation while yielding the same result. Following {\color{black}a} similar approach, by evaluating other elements of (\ref{Fisher_I}) using (\ref{derivations}) and the Parseval theorem, the Fisher matrix for a single path ($P=1$) is derived as (\ref{I_og}), shown on the top of this page. Finally, based on the i.i.d. path assumption and the block-diagonal property of the FIM, the CRLBs are obtained by {\color{black}inverting} $\mathbf{I}(\boldsymbol{\theta}, \boldsymbol{x})$ as:
\setcounter{equation}{35}
\begin{subequations}
\label{crlb}
\begin{equation}
{\rm CRLB}(|{\tilde h}_i|) = \frac{{\sigma_{\bf v}^2}}{2 NM},
\label{crlb_h}
\end{equation}
\begin{equation}
{\rm CRLB}(\phi_i) = \frac{{\sigma_{\bf v}^2}}{2\left|{\tilde h}_i\right|^2}{\frac{{7NM + N{d_f} + M{d_t} - 5{d_t}{d_f}}}{{MN\left( {N + {d_t}} \right)\left( {M + {d_f}} \right)}}},
\label{crlb_phi}
\end{equation}
\begin{equation}
{\rm CRLB}(k_i) = \frac{{\sigma_{\bf v}^2}}{2\left|{\tilde h}_i\right|^2}{\frac{{3N}}{{{\pi ^2}M\left( {{N^2} - d_t^2} \right)}}},
\label{crlb_k}
\end{equation}
\begin{equation}
{\rm CRLB}(l_i) =\frac{{\sigma_{\bf v}^2}}{2\left|{\tilde h}_i\right|^2} {\frac{{3M}}{{{\pi ^2}N\left( {{M^2} - d_f^2} \right)}}}.
\label{crlb_l}
\end{equation}
\end{subequations}

To this end, the unbiased estimation MSE lower bound of DD domain channel parameters in the DD-a-OFDM system are obtained.
\begin{remark}
Comparing our derived CRLBs in (\ref{crlb_k}) and (\ref{crlb_l}) with the OFDM radar bounds (see equations 3.56 and 3.57 of  \cite{Braun}), we observe exact correspondence when $d_f = 1$ and $d_t=1$. This consistency: (1) validates our pilot pattern design in Section \ref{pilotpattern}, and (2) confirms the correctness of (\ref{crlb}). Notably, while \cite{Braun} neglects TF domain ICI through the assumption $\Delta f \gg \nu_{\rm max}$, our analysis in Section \ref{ICI} demonstrates how this interference transforms into Gaussian noise in the DD domain, i.e. reflected by ${{\sigma_{\bf v}^2}}$, enabling more realistic CRLB derivation for practical systems.
\end{remark}

\section{DD domain Signal Processing-Aided Channel Estimation and Equalization}
Based on the designs of DD-a-OFDM and the principle of CSF estimation based on discrete pilots in Section \ref{section III}, we will further design the CSF estimators and equalizers for the DD-a-OFDM system in this section.
\subsection{Precise CSF Parameters Estimation}
Given the CRLB (\ref{crlb}) of the CSF parameters in the DD-a-OFDM system, it is worth performing accurate CSF estimation with the goal of approximating the CRLB. In the current DDMC-related research, there have been different strategies for CSF estimation \cite{add2,shi,jiyang}, which could be feasible for CSF estimation according to (\ref{h_DD}). Among which, we will introduce the maximum likelihood (ML) estimator to the DD-a-OFDM system.
\subsubsection{ML-based CSF Estimator}
\label{SECTIONXXXX}
According to the ML principle, {\color{black}a} ML estimator for (\ref{h_DD}) can be formulated based on (\ref{llf}) as:
\begin{equation}
\hat{\boldsymbol{\theta}}=\arg \min _{\boldsymbol{\theta} \in \mathbb{C}^{\hat P} \times \mathbb{R}^{\hat P} \times \mathbb{R}^{\hat P}} L(\hat h_{{\text{DD}}}^{{\text{Periodic}}} \mid \boldsymbol{\theta}),
\label{ML_estimator}
\end{equation}
where $\boldsymbol{\theta} \in {\mathbb C}^{1 \times 3{\hat P}}$ consists of ${\tilde h}_i$, ${l}_i$, and ${k}_i$ for $i = 1, \dots, {\hat P}$. Here, $\hat P$ denotes the number of estimated multipath components. Specifically, an element of $\hat h^{\rm Periodic}_{\rm DD}[k,l]$ is considered a valid multipath component contributing to $\hat{P}$ if $|\hat h^{\rm Periodic}_{\rm DD}[k,l]|^2 >{\cal E}_{\rm th}$, where ${\cal E}_{\rm th}$ is a threshold typically set to $3\sigma_{\bf v}$ \cite{Raviteja,zhiqiangoff}. Furthermore, $L(\hat h_{{\text{DD}}}^{{\text{Periodic}}} \mid \boldsymbol{\theta})$ represents the log-likelihood function to be minimized, which is defined according to (\ref{h_DD}) as:
\begin{equation}
L(\hat h_{{\text{DD}}}^{{\text{Periodic}}} \mid \boldsymbol{\theta})={\| {\hat {\bf{h}}_{{\text{DD}}}^{{\text{Periodic}}} - \sum\limits_{i = 1}^{\hat P} {{{\tilde h}_i}} {{\bf{\Phi }}_i}} \|^2},
\label{l1}
\end{equation}
where $\hat {\bf{h}}_{{\text{DD}}}^{{\text{Periodic}}} \in {\mathbb C}^{\frac{MN}{d_td_f} \times 1}$ refers to vectorizing $\hat h_{{\text{DD}}}^{{\text{Periodic}}}[k,l]$ for $k\in \{0,1,\frac{N}{d_t}-1\}$ and $l \in \{0,1,...,\frac{M}{d_f}-1\}$, with $\hat {\bf{h}}_{{\text{DD}}}^{{\text{Periodic}}}\left[k\frac{M}{d_f}+l\right] = \hat h_{{\text{DD}}}^{{\text{Periodic}}}[k,l]$. Besides, ${\bf \Phi}_i \in {\mathbb C}^{\frac{MN}{d_td_f} \times 1}$ refers to the basis vector associated with the $i$-th multipath, which is given by:
\begin{equation}
{\bf \Phi}_i[kM/d_f+l] = {\cal R}_{{\text{delay}}}^{{\text{Periodic}}}({l_i},l){\cal R}_{{\text{Doppler}}}^{{\text{Periodic}}}({k_i},k).
\end{equation}

Under the assumption that $l_i$ and $k_i$ are fixed, the minimization of (\ref{l1}) w.r.t. $\tilde h_i$ for fixed $l_i$ and $k_i$ can be obtained by the solution of the following equation:
\begin{equation}
\sum\limits_{j= 1}^{\hat P} {{{\tilde h}_j}} {{\bf{\Phi }}^H_i}{{\bf{\Phi }}_j} =\boldsymbol{\Phi}_i^H \hat {\mathbf{h}}_{\text{DD}}^{\text{Periodic}}.
\label{h_rel}
\end{equation}

Based on (\ref{h_rel}), one can suppose that there is not inter-path interference (IPI), i.e., ${\bf{\Phi }}_i^H{{\bf{\Phi }}_{j}}=0$ for $i \ne j$, and the ML estimator can be easily solved \cite{Gaudio}. Nevertheless, the basis vectors $\{{\bf \Phi}_i\}$ may not be strictly orthogonal under the off-grid channel conditions. Specifically, we note that for given $\{l_i, k_i\}$, the ML estimate of the complex gains $\tilde{\mathbf{h}} = [\tilde{h}_1, \dots, \tilde{h}_{\hat P}]^T$ can be given by the LS solution of (\ref{h_rel}) as:

\begin{algorithm}[t]
\caption{ML-based CSF estimator for DD-a-OFDM}
\label{alg1}
\begin{algorithmic}[1]
\REQUIRE $\hat P$, $\hat {\bf{h}}_{{\rm{DD}}}^{{\rm{Periodic}}}$, and iteration number ${\rm Num}_{\rm Iter}$.
\ENSURE $\hat {\boldsymbol{\theta}} = \{ \hat{\tilde h}_i, \hat l_i, \hat k_i \}_{i=1}^{\hat P}$.

\STATE \textbf{Initialization:} Initialize $\{ \hat l_i^{(0)}, \hat k_i^{(0)}, \hat{\tilde h}_i^{(0)},{\bf \Phi}^{(0)}\}$.
\FOR{$t = 1$ to ${\rm Num}_{\rm iter}$}
    \FOR{$i = 1$ to $\hat P$}
        \STATE \textbf{Residual Calculation:} Remove the interference of other paths:
        \begin{equation*}
            {\bf r}_i^{(t)} = \hat {\bf{h}}_{{\rm{DD}}}^{{\rm{Periodic}}} - \sum_{j < i} \hat{\tilde h}_j^{(t)} {\bf \Phi}_j^{(t)} - \sum_{j > i} \hat{\tilde h}_j^{(t-1)} {\bf \Phi}_j^{(t-1)}.
        \end{equation*}
        \STATE \textbf{Path-wise ML:} Update $(\hat l_i^{(t)}, \hat k_i^{(t)})$ by maximizing the compressed likelihood of the $i$-th path in the residual domain:
        \begin{equation*}
            (\hat l_i^{(t)}, \hat k_i^{(t)}) = \arg \max_{l_i, k_i} \frac{|{\bf \Phi}_i^H(l_i, k_i) {\bf r}_i^{(t)}|^2}{\|{\bf \Phi}_i(l_i, k_i)\|^2}.
        \label{step_compressed}
        \end{equation*}
        \STATE Update the $i$-th column of ${\bf \Phi}^{(t)}$ with ${\bf \Phi}_i(\hat l_i^{(t)}, \hat k_i^{(t)})$.
    \ENDFOR
    \STATE \textbf{Joint Gain Refinement:} Re-estimate all gains using (\ref{joint_ls}) to ensure global convergence.
\ENDFOR
\RETURN $\hat {\boldsymbol{\theta}}$
\end{algorithmic}
\end{algorithm}

\begin{equation}
\hat{\tilde{\mathbf{h}}} = ( \mathbf{\Phi}^H \mathbf{\Phi} )^{-1} \mathbf{\Phi}^H \hat{\mathbf{h}}_{\text{DD}}^{\text{Periodic}},
\label{joint_ls}
\end{equation}
where $\mathbf{\Phi} = [\mathbf{\Phi}_1, \dots, \mathbf{\Phi}_{\hat P}] \in {\mathbb C}^{\frac{MN}{d_td_f} \times \hat P}$. By substituting (\ref{joint_ls}) back into (\ref{ML_estimator}), the cost function becomes independent of the complex gains $\tilde{\mathbf{h}}$. This leads to the \textit{compressed likelihood function} $L_C(\boldsymbol{l}, \boldsymbol{k})$, which only depends on the non-linear parameters $\boldsymbol{l}$ and $\boldsymbol{k}$ and needs to be maximized:
\begin{equation}
L_C(\boldsymbol{l}, \boldsymbol{k}) = \| \mathbf{P}_{\boldsymbol{\Phi}}(\boldsymbol{l}, \boldsymbol{k}) \hat{\mathbf{h}}_{\text{DD}}^{\text{Periodic}} \|^2,
\label{compressed_LLF}
\end{equation}
where $\mathbf{P}_{\boldsymbol{\Phi}} = \mathbf{\Phi}(\mathbf{\Phi}^H\mathbf{\Phi})^{-1}\mathbf{\Phi}^H$ is the projection matrix onto the subspace spanned by the basis vectors, $\boldsymbol{l} = [{l}_1, \dots, {l}_{\hat P}]^T$, $\boldsymbol{k} = [{k}_1, \dots, {k}_{\hat P}]^T$. {\color{black}Note that directly maximizing (\ref{compressed_LLF}) requires a multi-dimensional search, which is computationally exhaustive. Thus, we adopt the \textit{Alternating Projection} (AP) approach to iteratively decompose the multi-dimensional optimization into a sequence of two-dimensional searches (i.e., joint delay and Doppler searches for each individual path). The detailed steps are shown in Algorithm \ref{alg1}.}\par


As summarized in Algorithm \ref{alg1}, we decompose the joint optimization into an iterative process. In each outer iteration $t$, we sequentially update each path's parameters by calculating the residual signal $\mathbf{r}_i^{(t)}$, which removes the interference from all other current path estimates. In Step 5, a refined search (e.g., golden section search or Newton's method) is performed on $\mathbf{r}_i^{(t)}$ to update $(\hat{l}_i, \hat{k}_i)$. Crucially, in Step 8, we perform a joint LS update for all complex gains. This step compensates for the non-orthogonality between paths, ensuring that the estimator converges to the joint ML solution.

\subsubsection{Complexity of Algorithms 1}
\label{comana}

In Algorithm \ref{alg1}, the computational complexity is primarily determined by the iterative process involving residual calculation, path-wise parameter search, and joint gain refinement. In Step 4, calculating the residual ${\bf r}_i^{(t)}$ for the $i$-th path requires ${\cal O}(\hat P \frac{MN}{d_td_f})$ operations. In Step 5, assuming the one-dimensional refinement involves ${\rm Num}_{\text{search}}$ steps, the complexity of calculating the normalized projections is ${\cal O}({\rm Num}_{\text{search}} \frac{MN}{d_td_f})$. In Step 8, the joint gain refinement requires constructing the correlation matrix ${\bf \Phi}^H{\bf \Phi}$ with ${\cal O}(\hat P^2 L)$ operations and performing a small-scale matrix inversion with ${\cal O}(\hat P^3)$ complexity. Considering the outer iteration number ${\rm { Num}}_{\text{iter}}$, the overall complexity of Algorithm \ref{alg1} is given by ${\cal O}\left( {\rm Num}_{\text{iter}} ( \hat P N_{\text{search}} \frac{MN}{d_td_f} + \hat P^2 \frac{MN}{d_td_f} + \hat P^3 ) \right)$. Since the number of paths is typically much smaller than the observation size, i.e., $\hat P \ll \frac{MN}{d_td_f}$, the complexity is approximately ${\cal O}(\hat P \frac{MN}{d_td_f} {\rm Num}_{\text{search}} {\rm Num}_{\text{iter}})$, which remains linear with respect to the pilot grid size $\frac{MN}{d_td_f}$.

Furthermore, to make smaller ${\rm Num_{search}}$ feasible for Algorithm 1, we can input coarse estimation of $\hat l_i$ and $\hat k_i$ as inputs for Algorithm \ref{alg1}, which can be obtained by the peak detection estimator in \cite{wlc}, rather than $\hat l_i=0, \hat k_i=0$. Afterwards, we can find the concise value of $\hat k_i$ based on the dichotomy or ternary methods for $\hat k_i \in [\hat k_i-\delta,\hat k_i+\delta]$ with $0<\delta \le 0.25$. Besides, Algorithm \ref{alg1} can also be terminated when the changes of $\hat {\boldsymbol{\theta}}$ is negligible, which is not shown for brevity.
\subsection{CTF Equalization}
Based on the estimated CSF parameters $\hat {\boldsymbol \theta}$, the estimation of TF domain channel matrix corresponding to the $n$-th OFDM symbol ${\hat {\bf H}}_{{\rm{TF}}}^n \in {\mathbb C}^{M\times M}$ can be directly derived according to (\ref{CSF_equ}). Afterwards, the TF domain multiplexed signal ${\bf x}_{\rm TF}^n$ can be detected after CTF equalization. Therein, conventional OFDM equalizers can be adopted, such as MMSE \cite{xuehan2} and MMSE based on successive interference cancellation which explores frequency domain diversity\cite{qi}. Besides, we note that the cross-domain (frequency and delay domains) equalizer designed based on DD domain signal processing principles could  compensate the DD-a-OFDM performance resulted from the non-Gaussian distribution of frequency domain symbols \cite{CD}. With emphasis on the DD domain signal processing aided CSF estimation and limited by space, in this article, the MMSE equalizer is considered. Based on the MMSE principle, the equalized TF domain symbols can be obtained as:
\begin{equation}
{\hat {\bf x}}^{n}_{\rm TF} = \mathbf{G}_{\mathrm{MMSE}}{{\bf y}}^{n}_{\rm TF},
\label{mmse}
\end{equation}
where $ \mathbf{G}_{\mathrm{MMSE}} \in {\mathbb c}^{M\times M}$ refers to the MMSE equalizer:
\begin{equation}
\mathbf{G}_{\mathrm{MMSE}}= ({\hat {\bf H}}^n_{\mathrm{TF}})^{{H}}\left({\hat {\bf H}}_{\mathrm{TF}}^n ({\hat {\bf H}}^n_{\mathrm{TF}})^{{H}}+\frac{\sigma_{\bf w}^2}{E_{\rm p}} \mathbf{I}_{M}\right)^{-1}.
\label{mmse_0}
\end{equation}
\begin{remark}
{\color{black}We note that in DD-a-OFDM, the estimation of CSF facilitates the precise acquisition of the ICI-impacted full-matrix TF channel ${\hat {\bf H}}_{{\rm{TF}}}^n \in {\mathbb C}^{M\times M}$, which is fundamentally difficult to obtain in conventional OFDM systems. This accurate full-matrix CSI is the key enabler for precise TF-domain ICI-aware equalization. Furthermore, under the DD-a-OFDM framework, the equalization and transmission designs are still open for exploration. For instance, exploiting the sparsity of the DD domain channel to design low-complexity cross-domain (e.g., delay-frequency) detectors, or improving the transmitter for enhanced Doppler diversity, represent promising future directions. Due to space constraints, this paper focuses on establishing the theoretical limits and optimal algorithms for DD domain channel estimation, leaving advanced DD-a-OFDM equalization designs for our future research.}
\end{remark}
\section{Numerical Results}
In this section, we evaluate the transmission performance of the proposed DD-a-OFDM scheme in high-mobility scenarios. Unless otherwise specified, all simulations employ the parameters listed in Table \ref{tab1}. We model the multipath components with uniformly distributed delays and Doppler shifts, i.e, $\tau_i \sim \mathcal{U}(0,\tau_{\text{max}})$ for delays and $\nu_i \sim \mathcal{U}(-\nu_{\text{max}},\nu_{\text{max}})$ for Doppler shifts, where all paths have equal power coefficients. Besides, we set delays of multipath as on-grid\footnote{{\color{black}Herein, we assume on-grid delay condition to provide a proof-of-concept for the proposed DD-a-OFDM framework. We acknowledge the limitations of this assumption when applied to realistic channel models like 3GPP TDL-C, which exhibit continuous power delay profiles and severe off-grid fractional delays. In future work, we will explicitly derive the off-grid DD-a-OFDM system model and its corresponding CSF estimation CRLB, design low-complexity ML channel estimators capable of resolving continuous clusters, and evaluate the performance of DD-a-OFDM under TDL-C and other practical channel conditions.}} w.r.t. $\frac{1}{M\Delta f}$ and Dopplers of multipath as off-grid w.r.t. $\frac{1}{NT}$.

\begin{table}[t]
\caption{Parameter Settings}
\centering
\begin{tabular}{c|c|c}
  \hline
  {\bf Physical meaning} & {\bf Parameters} & {\bf Values} \\
  \hline
  {\color{black}Carrier frequency (GHz)} & - & 2.1\\
  Number of subcarriers & $M$ & 64\\
  Number of OFDM symbols & $N$ & 64\\
  Pilot spacing in frequency domain & $d_f$ & 4\\
  Pilot spacing in time domain & $d_t$ & 4\\
  Subcarrier spacing (kHz) & $\Delta f$& 15\\
  Duration of one OFDM symbol ($\mu$s) & $T$& 66.7\\  
  Number of multipath & $P$ & 5\\
  Maximum velocity (m/s) & $-$ & 100 \\
  Maximum delay ($\mu$s) & $ \tau_{\rm max}$ & 4.17\\
  Maximum Doppler shift (Hz) & $ \nu_{\rm max}$& 937.5\\
  Modulation Type & $-$ & 4-QAM\\
  \hline
\end{tabular}
\label{tab1}
\end{table}
\subsection{Benchmark Schemes}
The conventional OFDM system with traditional CTF estimators is selected as benchmark, whose principle is briefly introduced below. In the conventional OFDM system, it is difficult to estimate ${\bf H}^{n}_{\rm TF} \in {\mathbb C}^{M\times M}$, since the dimension {of ${\bf H}^{n}_{\rm TF}$ exceeds both the received signal dimension (${\bf y}^{n}_{\rm TF} \in {\mathbb C}^{M\times 1}$) and the available pilots (${\bf x}^{n}_{\rm TF}[m'd_f,n'd_t]$ with $m' \in \{0,1,..,\frac{M}{d_f}\}$ and $n' \in \{0,1,..,\frac{N}{d_t}\}$).} Therefore, the conventional OFDM typically ignores ICI by assuming $\nu_{\rm max} \ll \Delta f$, under which condition the I/O relationship in (\ref{OFDMIOTD}) can be approximated as \cite{Gaudio}:
\begin{equation}
{\bf y}_{\rm TF}[m,n] \approx \sum^{P}_{i = 1} h_i {e^{j2\pi \frac{nk_i}{N}}}{{e^{\frac{{ - j2\pi m{l_i}}}{M}}}}{\bf x}_{\rm TF}[m,n] + {\bf w}.
\label{OFDMIOTDapp}
\end{equation}

Without loss of generality, the CTF associated with pilots will be estimated based on the LS principle same as (\ref{TF_LS}), to obtain ${\hat h}^{\rm Pilot}_{\rm TF}[m,n]$. Afterwards, two typical estimators will be used, including the linear interpolation estimator and the MMSE estimator.
\subsubsection{Linear Interpolation Estimation of CTF}
Based on the principle of linear interpolation, the time domain interpolated CTF can be obtained as:
\begin{equation}
\small
\begin{aligned}
\!\!{\hat h}_{\rm TF}[m,n] & \!=\!({\hat h}^{\rm Pilot}_{\rm TF}[m,(\tilde n+1)d_t] \!-\!{\hat h}^{\rm Pilot}_{\rm TF}[m,\tilde nd_t]) {n'}/{d_t}\\
& +{\hat h}^{\rm Pilot}_{\rm TF}[m,\tilde nd_t],
\end{aligned}
\label{LI_cla}
\end{equation}
where $\tilde n = \lfloor \frac{n}{d_t} \rfloor$, $n' = {\rm mod}(n,{d_t})$. After performing linear interpolation along the frequency domain similarly, the CTF corresponding to $NM$ REs is obtained.
\subsubsection{MMSE Estimation of CTF}
Based on the principle of MMSE estimation, the estimated CTF will be obtained as \cite{b3}:
\begin{equation}
\begin{aligned}
\!{{\rm vec}(\hat{h}_{\rm TF})}\!=\!\mathbf{R}_{1}\!(\!\mathbf{R}_{2}\!+\!\sigma_{\bf w}^2 {\mathbb E}\left(\mathbf{x}_{\rm TF} \mathbf{x}_{\rm TF}^{\mathrm{H}}\right)^{-1}\!)\!^{-1}\!{\rm vec}({\hat h}^{\rm Pilot}_{\rm TF}),
\end{aligned}
\label{MMSE_cla}
\end{equation}
where ${\rm vec}(\cdot)$ denotes the vectorizing operation, $\mathbf{R}_{1} = {\mathbb E}({{\rm vec}(\hat{h}_{\rm TF})}{{\rm vec}(\hat{h}_{\rm Pilot})^H})$, and $\mathbf{R}_{2} = {\mathbb E}({{\rm vec}(\hat{h}^{\rm Pilot}_{\rm TF})}{{\rm vec}(\hat{h}_{\rm TF}^{\rm Pilot})^H})$. Notably, the MMSE estimator for the conventional OFDM system requires the exact statical information of CTF, and whose complexity is ${\cal O}(\frac{M^3}{d_f^3}\frac{N^3}{d^3_t})$.\par
\subsection{\color{black}{Validation of CRLB for CSF Estimation in DD-a-OFDM}}
\begin{figure}[htbp]
  \centering
  \includegraphics[width=.495\textwidth]{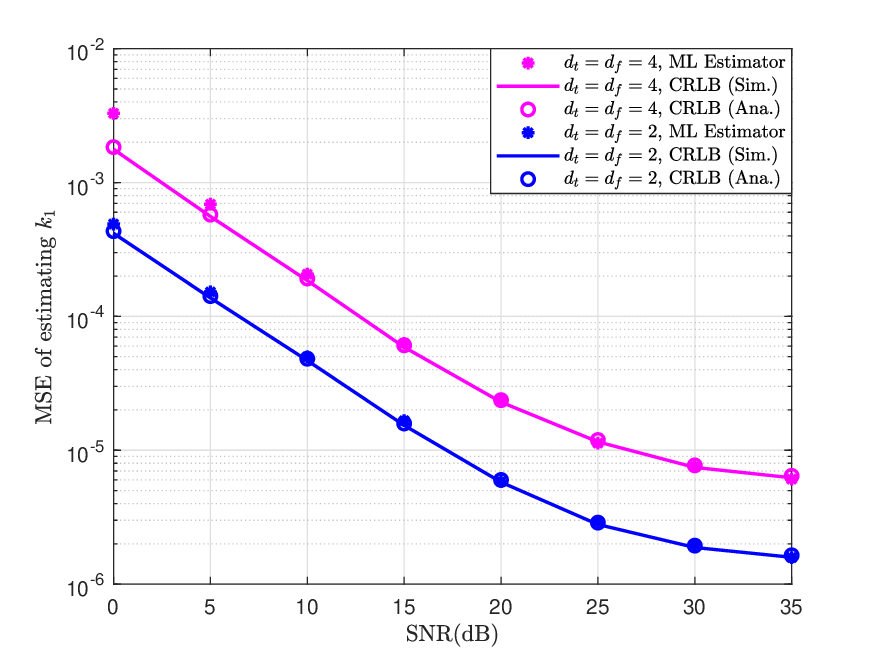}
  \caption{\color{black}MSE performance of $k_1$ in the DD-a-OFDM system.}\label{FIG4}
  \vspace{-1em}
\end{figure}
\begin{figure}[htbp]
     \centering
     \subfigure[]{\label{fig:sub1}\includegraphics[width=0.45\textwidth]{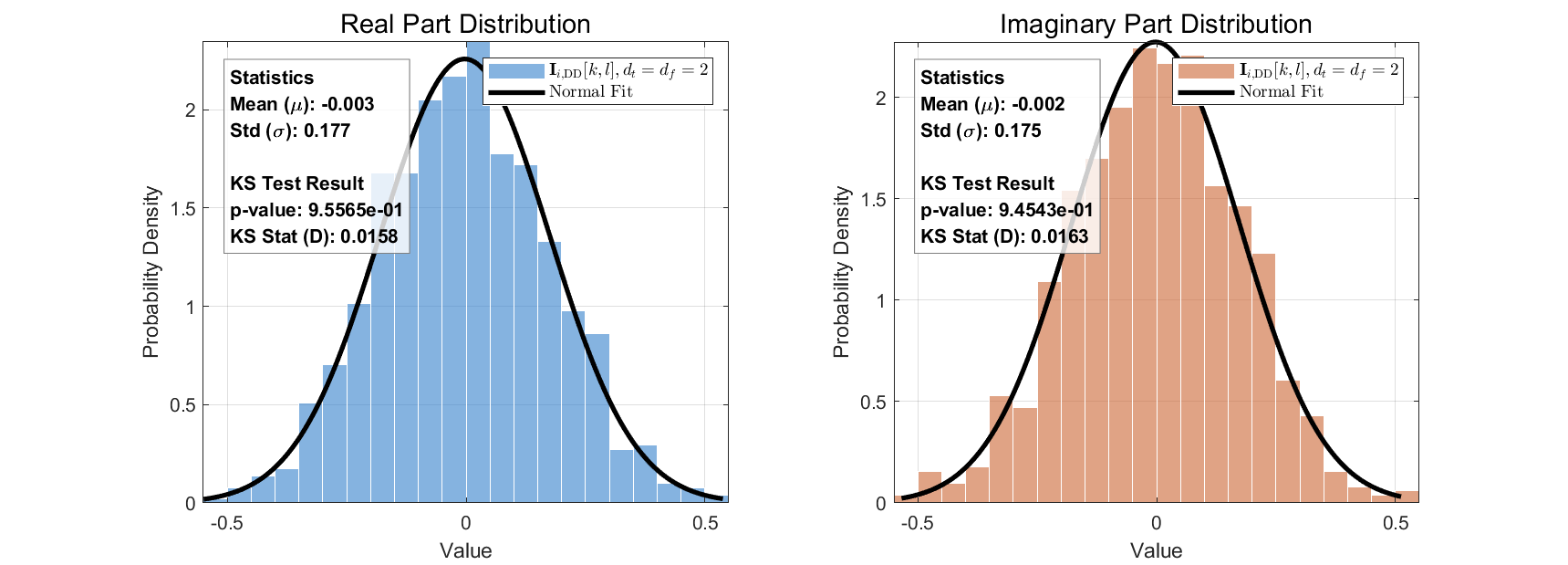}}
     \vspace{-1em}
    \hfill
    \subfigure[]{\label{fig:sub2}\includegraphics[width=0.45\textwidth]{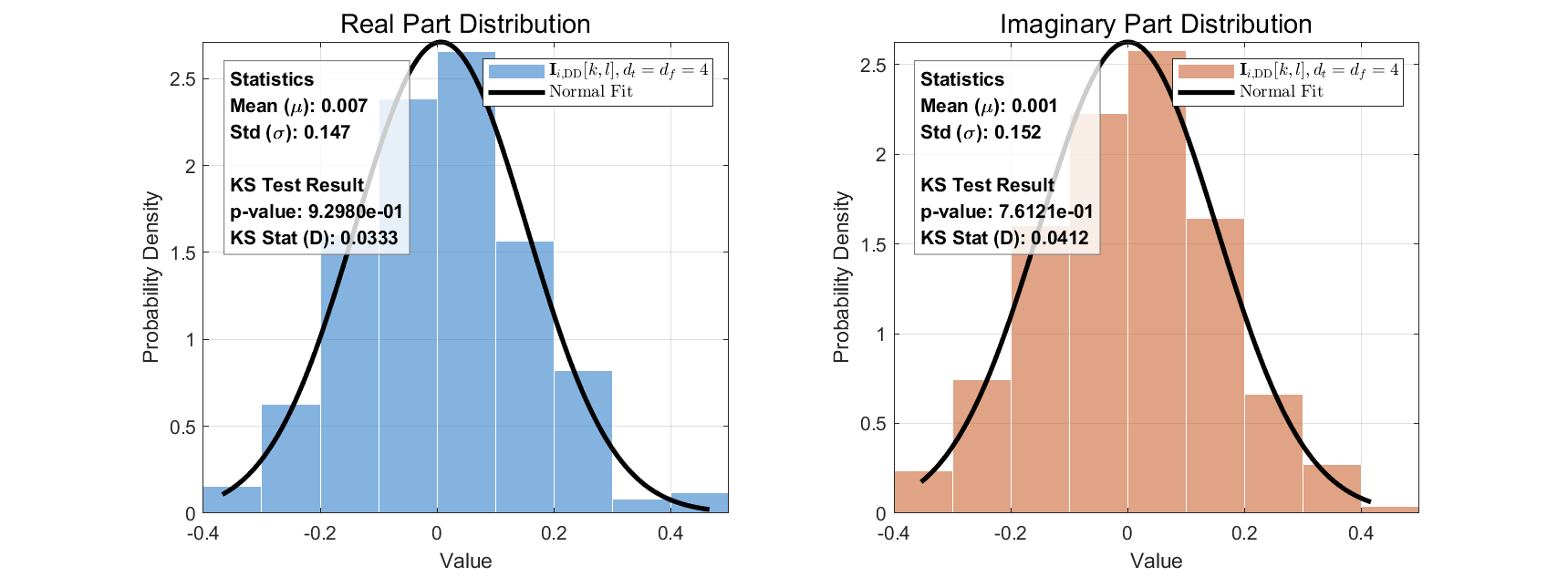}}
     \caption{Distribution fitting of DD domain interference ${\bf I}_{i,{\rm DD}}[k,l]$ in (\ref{I_DD}) {\color{black}with} (a) $N = 64$, (b) $N = 128$.}
     \vspace{-1em}
     \label{FIG4A}
\end{figure}
\begin{figure}
  \centering
  \includegraphics[width=.495\textwidth]{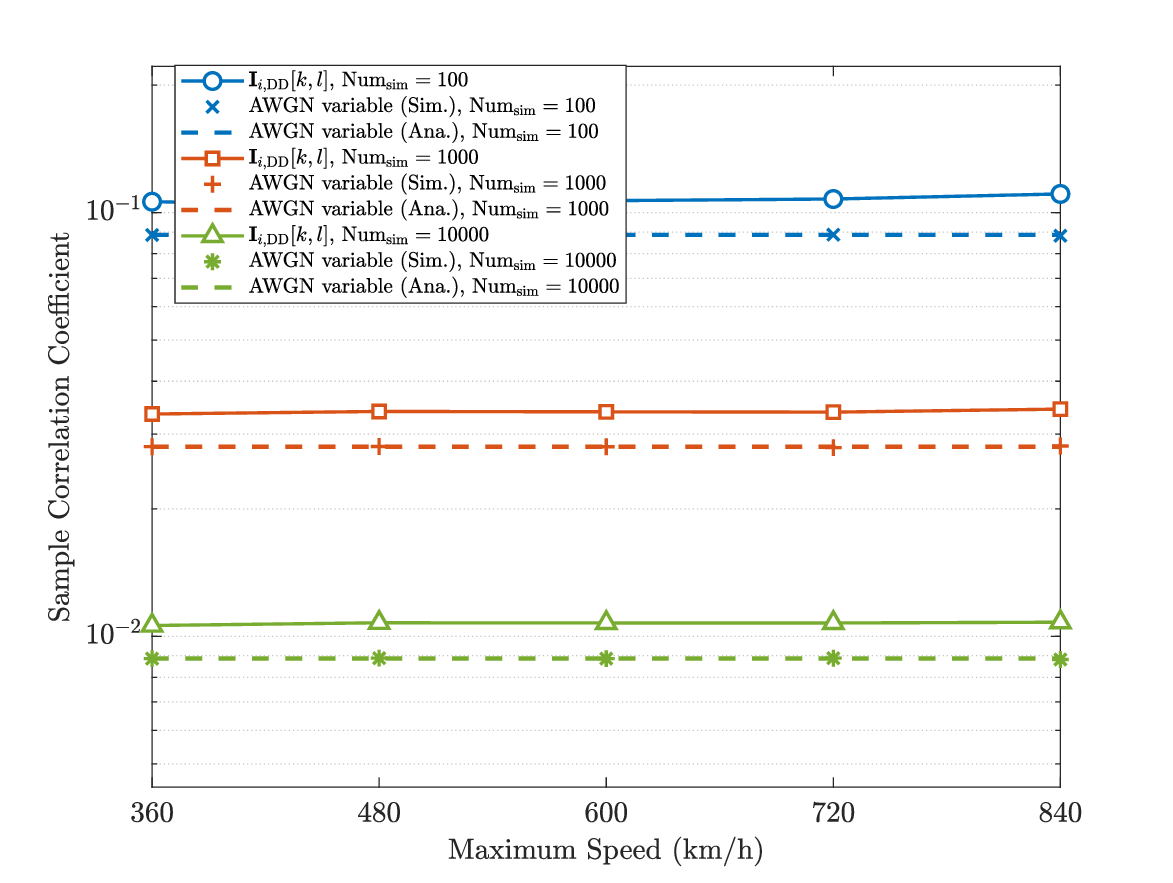}
  \caption{{\color{black}Sample correlation coefficient of ${\bf I}_{i,{\rm DD}}[k,l]$ in (\ref{I_DD}).}}\label{FIG5A}
  \vspace{-1em}
\end{figure}
In the following, we first evaluate the accuracy of the proposed receiving strategies and the derived CRLB for CSF estimation in the DD-a-OFDM system. Based on the identical multipath power settings, Fig. \ref{FIG4} plots the MSEs of $k_1$ for the ML estimator (Algorithm \ref{alg1}), alongside {\color{black} the exact simulated CRLB considering IPI based on \eqref{Fisher_I_full} and \eqref{crlb111} (denoted as CRLB (Sim.)), and the analytical CRLB derived in (\ref{crlb_k}) ignoring IPI (denoted as CRLB (Ana.))} for $d_t=d_f=4$ and $d_t=d_f=2$ scenarios. The MSEs of $|h_i|$, $\phi_i$, $l_i$ in (\ref{crlb_h}), (\ref{crlb_phi}) and (\ref{crlb_l}) exhibit similar trends to $k_1$ and are omitted for brevity.\par

First, Fig. \ref{FIG4} demonstrates that {\color{black} the simulated exact CRLB and the analytical approximated CRLB are virtually indistinguishable across the entire SNR regime. This objectively confirms that the IPI coupling introduces a negligible impact on the CRLB, validating the rationality of the block-diagonal FIM approximation in (\ref{Fisher_I}). Furthermore,} the ML estimator's MSE approaches the CRLB at both $M/d_f=N/d_t=16$ and $M/d_f=N/d_t=32$ in the high-SNR regime, validating both the CRLB derivation in (\ref{crlb_k}) and the feasibility of Algorithm \ref{alg1} {\color{black} which explicitly addresses the IPI in practice}. Besides, we set ${\rm Num}_{\text {iter}} = 3$ and ${\rm Num}_{\text {search}} = 8$ for the ML estimator in simulations. Consequently, the complexity of the ML estimator is ${\cal O}(24\hat P\frac{MN}{d_td_f})$. Furthermore, Fig. \ref{FIG4} shows that the CRLB curves reach MSE floors at high SNRs, confirming the analysis in (\ref{h_DD}). This floor is attributed to the equivalent Gaussian interference arising from ICI. As AWGN becomes negligible, the TF domain ICI from data symbols ultimately constrains the CSF estimation accuracy. While not the primary focus of this work, ICI mitigation is addressable by incorporating feedback loops between CSF estimation and channel equalization \cite{BEM1}, a topic left for future study.\par



{\color{black} Within the CRLB in (\ref{crlb_k}), the i.i.d. Complex Gaussian assumption of ${\bf I}_{i,{\rm DD}}[k,l]$ in (\ref{I_DD}) is concerning.} For visualization, the distribution of ${\bf I}_{i,{\rm DD}}[k,l]$ in (\ref{I_DD}) and its corresponding fitting results are presented in Fig. \ref{FIG4A}. It can be observed that increased $N/d_t$ and $M/d_f$ lead to superior Kolmogorov-Smirnov (K-S) test results. The results in Fig. \ref{FIG4A} confirm that the Gaussian distribution assumption for ${\bf I}_{i,{\rm DD}}[k,l]$ in (\ref{I_DD}) remains valid when $N/d_t \ge 16$ and $M/d_f \ge 16$. {\color{black} Furthermore, to validate the independence assumption of ${\bf I}_{i,{\rm DD}}[k,l]$ in (\ref{I_DD}), we plot the sample correlation coefficient of ${\bf I}_{i,{\rm DD}}[k,l]$ in Fig. \ref{FIG5A} with $M=N=64$ and $d_t=d_f =4$. Therein, the sample correlation coefficient is defined as ${\mathbb E}\{{\bf I}_{i,{\rm DD}}[k,l]{\bf I}^*_{i,{\rm DD}}[k',l']\}/{\mathbb E}\{{\bf I}_{i,{\rm DD}}[k,l]{\bf I}^*_{i,{\rm DD}}[k,l]\}$ with $k'\ne k$ and $l'\ne l$. As the number of Monte Carlo simulations $\rm Num_{ sim}$ increases, the correlation coefficient tightly tracks the theoretical correlation coefficient of purely independent complex Gaussian variables ($\sqrt{\pi}/(2\sqrt{\rm Num_{sim}})$ \cite{bxxx}). Notably, the correlation factor remains consistently around 1.2 times that of ideal AWGN and shows no significant increase as the user speed rises from 360 km/h to 840 km/h. This confirms that the weak local correlation is dominated by the static Dirichlet-kernel structure of ${\bf A}_i$ in (\ref{A_11}), thereby justifying the treatment of ${\bf I}_{i,{\rm DD}}[k,l]$ as an i.i.d. Complex Gaussian variable for the CRLB derivation.}\par

\subsection{Transmission Performance Compared to the Conventional OFDM System}


\begin{figure}
  \centering
  \includegraphics[width=.495\textwidth]{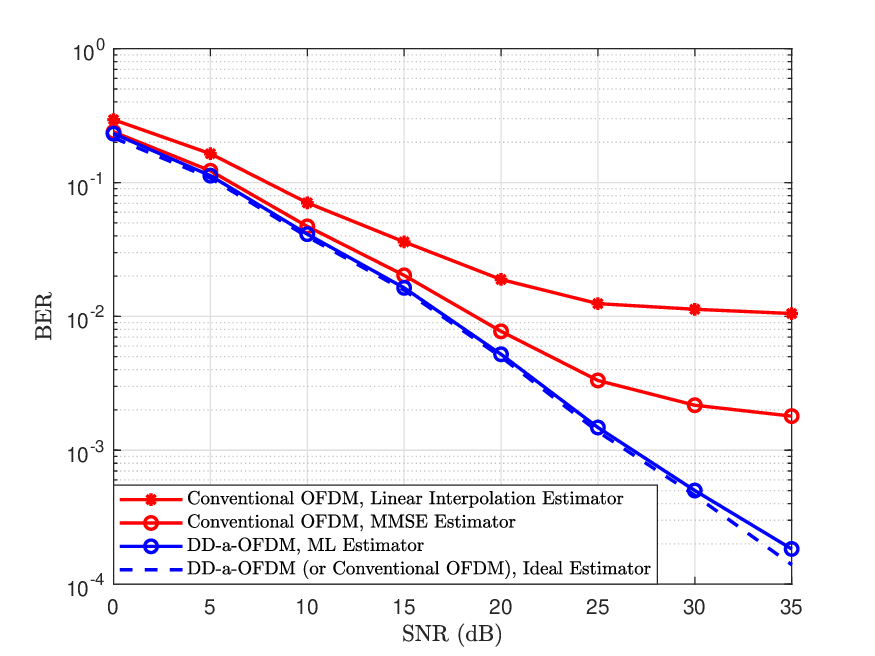}
  \caption{BER performance of DD-a-OFDM compared to conventional OFDM.}\label{FIG5}
  \vspace{-1em}
\end{figure}

Having validated the theoretical derivations and algorithms for DD-a-OFDM, we evaluate its BER performance against a conventional OFDM system, as illustrated in Fig. \ref{FIG5}. To ensure a fair comparison, all schemes share identical simulation settings and employ an MMSE equalizer, differing only in their channel estimation methods. Specifically, the proposed designs are compared with the MMSE estimator in \eqref{MMSE_cla} and the linear interpolation estimator in \eqref{LI_cla}.\par

As illustrated in Fig. \ref{FIG5}, several key observations can be made. First, the DD-a-OFDM system consistently outperforms the conventional OFDM system in high-mobility environments, particularly in the high-SNR regime. Specifically, the conventional OFDM system utilizing a linear interpolation estimator encounters an error floor at high SNRs, primarily due to unmitigated ICI and the estimator's inability to capture rapid CTF variations. While the MMSE-based OFDM system matches the performance of our proposed design at low SNRs, where the BER is noise-limited, it also suffers from an error floor in the high-SNR region. In contrast, the DD-a-OFDM system with the ML estimator closely approaches the ideal performance bound. This demonstrates that the proposed system effectively manages ICI through DD domain CSF estimation, overcoming a critical bottleneck of conventional OFDM in high-mobility scenarios.\par

{\color{black}Furthermore, the total receiver complexity is analyzed below. For the estimation stage, the traditional full TF-domain MMSE estimator entails a complexity of $\mathcal{O}((\frac{MN}{d_f d_t})^3)$, which exceeds that of the proposed ML estimator. As derived based on Algorithm 1, our estimation complexity is $\mathcal{O}(\hat P \frac{MN}{d_td_f} {\rm Num}_{\text{search}} {\rm Num}_{\text{iter}})$. Regarding the equalization stage, both systems in this work utilize full-matrix MMSE with $\mathcal{O}(N M^3)$ to provide a performance upper bound. Consequently, the current total complexity of DD-a-OFDM is $\mathcal{O}(\hat P \frac{MN}{d_td_f} {\rm Num}_{\text{search}} {\rm Num}_{\text{iter}} + NM^3)$, while that of the classical TF-domain MMSE system is $\mathcal{O}((\frac{MN}{d_f d_t})^3 + NM^3)$. Note that in future work, by integrating the proposed ML estimator with the cross-domain equalizer in [42], the total complexity of DD-a-OFDM is expected to be reduced to approximately $\mathcal{O}(\hat P \frac{MN}{d_td_f} {\rm Num}_{\text{search}} {\rm Num}_{\text{iter}} + NM \log M)$.}


Moreover, even with the ML estimator, the BER performance does not perfectly match the ideal bound at high SNRs. This phenomenon is attributed to the impact of ICI on DD domain CSF estimation in (\ref{h_DD}), which is consistent with the nonlinear CRLB curves observed in Fig. \ref{FIG4}. These observations motivate further research into mitigating the impact of Gaussian-distributed DD domain ICI.\par

\subsection{Transmission Performance under Different {\color{black}Speed} Conditions}
\begin{figure}
\centering
\subfigure[]{
\includegraphics[width=.495\textwidth]{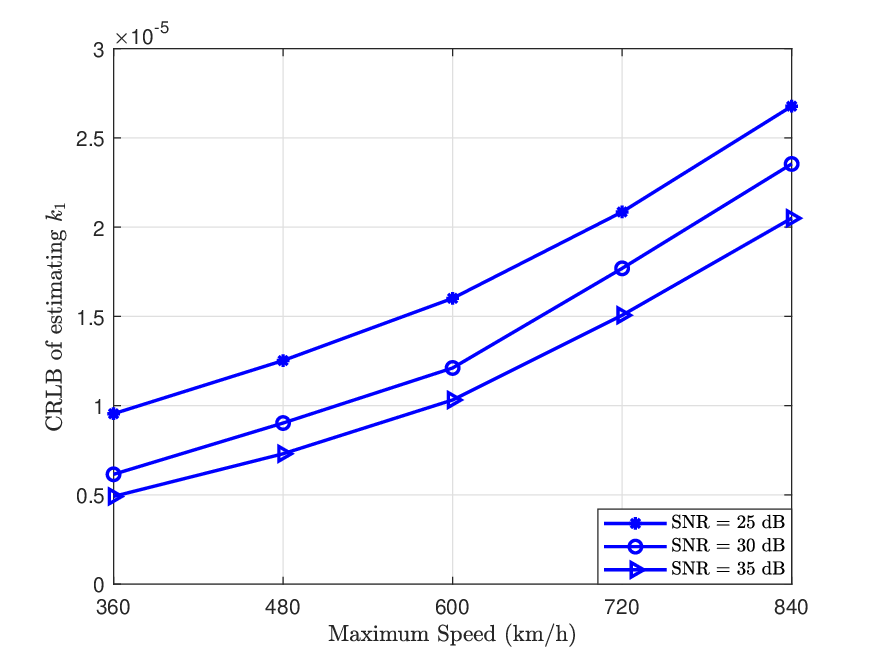}
\label{FIG6a}
}
\vspace{-1em}
\subfigure[]{
\includegraphics[width=.495\textwidth]{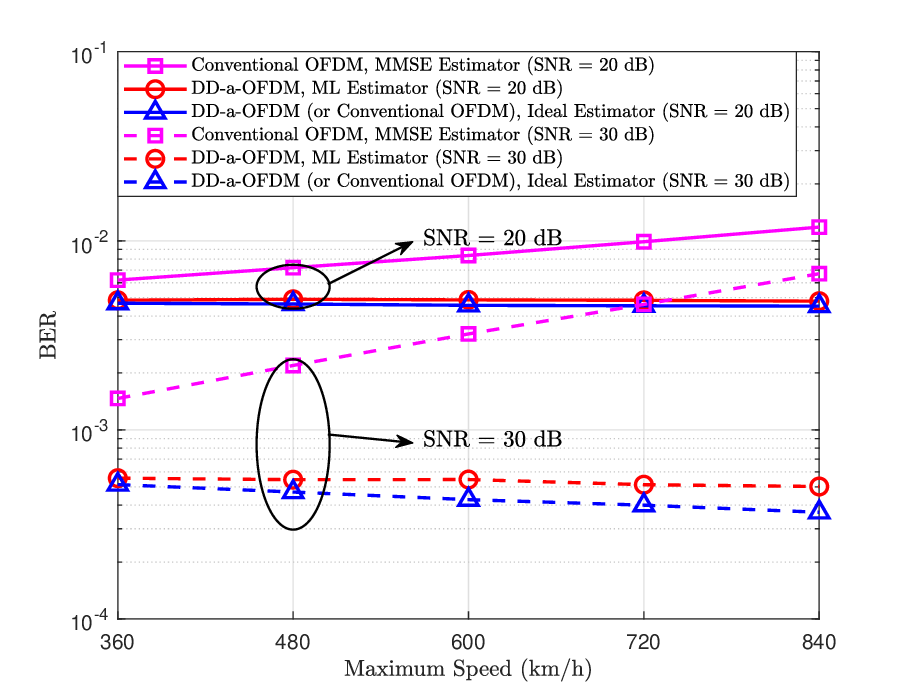}
\label{FIG6b}
}
\label{FIG6xx}
\caption{{\color{black}Performance of DD-a-OFDM under different Speed conditions: (a) MSE and (b) BER.}}
\vspace{-1em}
\end{figure}

In this section, we evaluate the MSE and BER performance of the proposed DD-a-OFDM system under a range of maximum Doppler conditions to verify its robustness in high-mobility scenarios. As illustrated in Fig. 8, {\color{black}the maximum speed varies from 360 km/h to 840 km/h, corresponding to the} maximum Doppler shift $\nu_{\max}$ varying from $703.125$ Hz to $1640.625$ Hz. This interval is specifically selected to satisfy the constraint $\nu_{\max} \in [0, \frac{1}{2d_tT}-\frac{1}{NT}]$, which, as stated in Section \ref{DDPE}, represents the fundamental limit for unambiguous CSF estimation using the proposed discrete pilot pattern.\par

Fig. \ref{FIG6a} illustrates the CRLB of the normalized Doppler index $k_1$ as a function of {\color{black}mobility speed} for different operating SNRs. It is evident that the CRLB exhibits a monotonically increasing trend with respect to the {\color{black} speed}, an observation that aligns with the theoretical modeling of the equivalent noise variance $\sigma^2_{\mathbf{v}}$ in \eqref{sigma_v} {\color{black}and CRLB in \eqref{crlb_k}}. From a physical perspective, a higher {\color{black} speed} enhances the power leakage between subcarriers in the TF domain. This leads to a corresponding rise in the ICI-related residual term $(1-|{\mathbf{A}}_i[0,0]|^2)E_{\rm p}\sigma^2_{1/x}d_t d_f$ in \eqref{sigma_v}, which elevates the equivalent noise floor $\sigma^2_{\mathbf{v}}$ in the DD domain. Consequently, {\color{black}the CRLB associated with the Doppler parameter $k_i$ increases in \eqref{crlb_k}}.\par


{\color{black}Furthermore, Fig. \ref{FIG6b} compares the BER of the proposed DD-a-OFDM system using the ML CSF estimator against the conventional OFDM system utilizing a standard TF domain MMSE channel estimator across varying user speeds at SNR = 20 dB and 30 dB. Both systems employ an ICI-aware MMSE equalizer to maintain a fair baseline for comparison.}
First, focusing on the ideal channel estimation benchmarks (represented by the solid and dashed blue curves with triangle markers), a slight BER improvement is observed in the high-SNR regime as the speed increases. {\color{black}This phenomenon is a direct result of the Doppler diversity provided by higher mobility \cite{Wu_Survey_2016}. As elucidated in \cite{Wu_Survey_2016}, a larger Doppler spread leads to a reduced channel coherence time. Consequently, the transmitted signal experiences a larger number of independent channel fading realizations within a single transmission block. When perfect CSI is available at the receiver, the ICI-aware MMSE equalizer can effectively resolve and collect the dispersed energy across these time-varying states, thereby transforming the rapid channel fluctuations into beneficial diversity gains rather than destructive interference.}\par

{\color{black}However, the performance of practical estimators tells a different story. For the conventional OFDM system (indicated by the solid and dashed magenta curves with square markers), the BER performance degrades continuously and severely as the speed scales up. Conventional TF domain MMSE estimators typically fail to track the fast-varying structured ICI induced by high mobility, treating it instead as unstructured interference. Notably, as the speed approaches $840$ km/h, the performance gap between the 30 dB and 20 dB SNR curves narrows significantly. This convergence implies that increasing the transmit power no longer yields a meaningful reduction in BER, indicating that the conventional OFDM receiver is bottlenecked by a severe ICI-induced error floor effect, which renders reliable communication nearly impossible.}\par

{\color{black}In sharp contrast, the proposed DD-a-OFDM system (represented by the solid and dashed red curves with circle markers) demonstrates robustness to varying mobility conditions. As the speed increases from $360$ km/h up to $840$ km/h, the BER curves remain highly stable and experience almost no improvement, tightly following the ideal estimation bounds. Even at extreme speeds, the DD-a-OFDM system avoids the severe performance deterioration seen in conventional OFDM. This confirms that by parameterizing and estimating the channel in the DD domain, the proposed framework effectively handles the time-varying nature of the channel and suppresses ICI more efficiently than traditional OFDM, supporting robust communications under extreme mobility.}

\subsection{Transmission Performance Compared to the OTFS System}

{\color{black}In this subsection, we conduct a comprehensive comparison between the proposed DD-a-OFDM system and the conventional OTFS system in \cite{Gaudio}. To maintain a fair baseline, the OTFS configuration adopts an embedded-pilot (EP) pattern where a single pilot symbol is placed at the center of the DD domain grid\footnote{\color{black}The rationale for adopting EP-OTFS as the baseline is to ensure a mathematically fair comparison under identical resource constraints. Since both schemes share structurally similar observation expressions in \eqref{h_DD}, they may achieve similar CRLB performance when subjected to the same observation overhead and free of data interference. While advanced designs such as scattered or superimposed pilots can further optimize OTFS overhead, they are not adopted here as they either restrict the supportable mobility range due to aliasing or introduce non-orthogonal pilot-data interference that contradicts our orthogonal multiplexing configuration. }, shielded by a guard interval of size $2\Delta k \times 2\Delta l$. Here, $\Delta k = N/(2d_t)$ and $\Delta l = M/(2d_f)$ denote the one-sided guard lengths in the Doppler and delay domains, respectively, ensuring that the pilot-occupied area in OTFS is equivalent to the pilot spacing in DD-a-OFDM. Furthermore, the OTFS pilot energy is boosted to $4\Delta k\Delta l$ to align the total power ratio between the two systems. Besides, Both evaluated schemes utilize the ML estimator initialized by peak detection outputs proposed in Section IV-A, and their theoretical CRLBs can be derived following the principles in \eqref{h_DD}.}

\begin{figure}
  \centering
  \includegraphics[width=.48\textwidth]{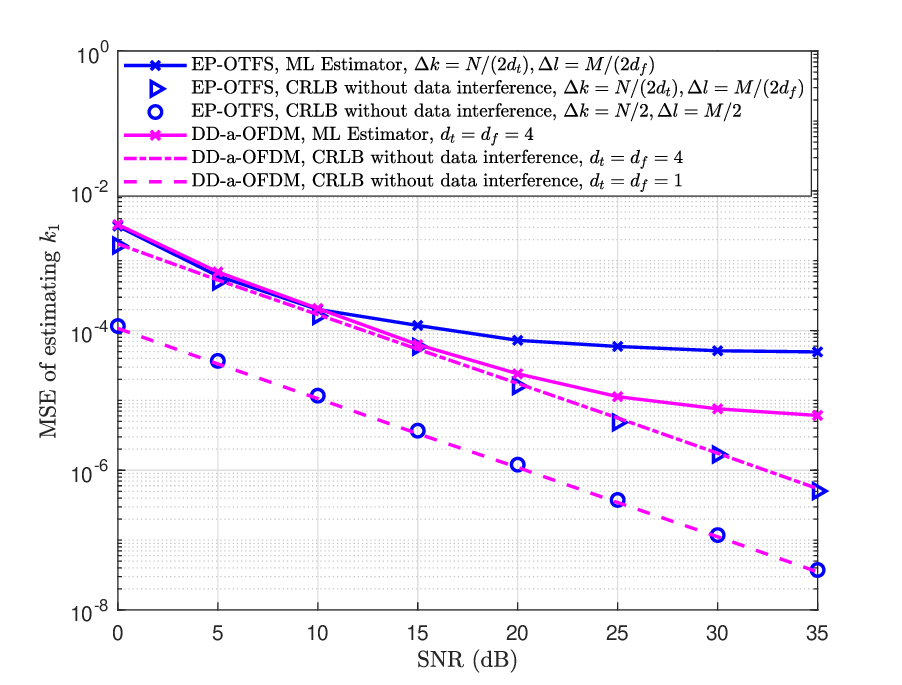}
  \caption{{\color{black}MSE performance of DD-a-OFDM system compared to OTFS.}}\label{FIG8A}
\end{figure}
{\color{black}Fig. \ref{FIG8A} illustrates the MSE of the normalized Doppler index $k_1$ estimation for both DD-a-OFDM and EP-OTFS systems. First, we plot the theoretical CRLB bounds without data interference. The interference-free CRLB for EP-OTFS is obtained by transmitting the pilot symbol without embedded data, whereas the CRLB for DD-a-OFDM is derived by setting the variance of the equivalent interference $\mathbf{I}_{i,{\rm DD}}$ to zero in \eqref{crlb_k}. In practice, such interference-free conditions in DD-a-OFDM can be approached by designing an iterative ICI cancellation mechanism between the channel estimator and the equalizer in future work. As demonstrated, under identical pilot overhead, average frame energy, and channel conditions, the CRLBs of DD-domain channel estimation for both systems are strictly equivalent in the absence of data interference. This is validated by the overlapping curves of the DD-a-OFDM CRLB (blue triangle and circle points) and the EP-OTFS CRLB (magenta dashed and dash-dot lines). This equivalence occurs because their DD domain observations share the same mathematical expression, as given in (\ref{h_DD}). Theoretically, this proves that purely utilizing TF domain pilots can achieve estimation performance comparable to that of DD-domain multiplexing schemes like EP-OTFS, which holds significant implications for OFDM-based high-mobility communications and integrated sensing and communications in 6G. \par

\begin{figure}
  \centering
  \includegraphics[width=.48\textwidth]{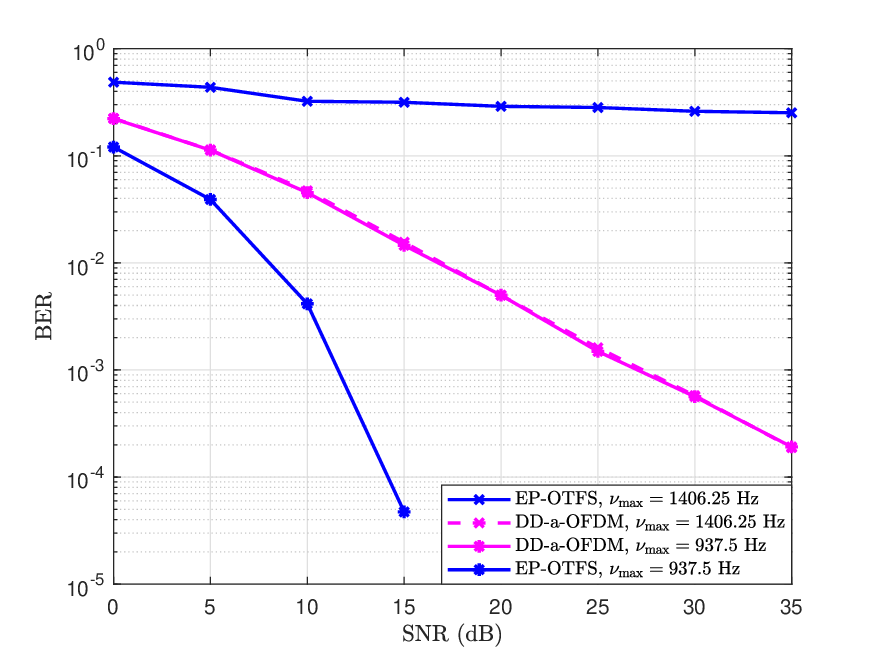}
  \caption{BER performance of DD-a-OFDM system compared to OTFS.}\label{FIG8B}
\end{figure}
Furthermore, we evaluate the practical ML estimation performance considering the impact of data interference. As shown by the MSE curves of the ML estimator for DD-a-OFDM at $d_t=d_f=4$ (magenta blue line with cross point) and EP-OTFS at $\Delta k=N/(2d_t), \Delta l=M/(2d_f)$ (solid blue line with cross point), DD-a-OFDM achieves significantly better channel estimation accuracy under the same pilot overhead constraint. This performance gap arises because the residual data interference in DD-a-OFDM behaves as an effective Gaussian process, which perfectly aligns with the proposed ML architecture. Conversely, the data-to-pilot interference in EP-OTFS, induced by Dirichlet sampling, exhibits severe non-Gaussian characteristics. This non-Gaussian interference significantly restricts the performance of standard ML estimators, rendering the low-complexity ML estimator proposed in Section IV-B suboptimal for EP-OTFS and leading to a pronounced error floor. \par}

{\color{black}Fig. \ref{FIG8B} compares the BER performance of the proposed DD-a-OFDM and EP-OTFS systems under different maximum Doppler shifts, which use MMSE equalizers. A primary observation is the performance degradation of EP-OTFS as $\nu_{\rm max}$ increases, e.g., $\nu_{\rm max}=1406.25$ Hz. This occurs because the stringent isolation requirement of DD domain channel estimation is violated in high-speed scenarios. To guarantee interference-free channel estimation, EP-OTFS would require one CP of length $\frac{M}{d_f}$ and $4\frac{N}{d_t}\frac{M}{d_f}$ REs for its pilot and guard interval. Meanwhile, the total overhead (including CPs and pilots) in the proposed DD-a-OFDM is $(N + \frac{N}{d_t}) \frac{M}{d_f}$ REs. However, this error floor could potentially be mitigated by designing channel estimators that explicitly account for data symbol interference. Even so, we can observe that by leveraging the periodic CSF property and the effectively Gaussian data interference, DD-a-OFDM could maintain reliable estimation performance with a lower overhead, overcoming the limitations of EP-OTFS in resource-constrained scenarios. Besides, DD-a-OFDM offers immediate hardware benefits by avoiding the high peak-to-average power ratio penalty associated with EP-OTFS's boosted pilots, as it utilizes TF domain pilots with energy levels comparable to data symbols.}

Nevertheless, it is objective to note that at a lower Doppler shift ($\nu_{\rm max} = 937.5$ Hz), OTFS achieves a superior BER floor compared to DD-a-OFDM in the high-SNR regime. This advantage stems from full-frame resource multiplexing in the DD domain, which allows OTFS to better harvest time-frequency domain channel diversity through MMSE equalization. It serves as a strong motivation for future research into DD-a-OFDM to capture similar diversity gains.

\section{Conclusion and Further {\color{black}Research}}

This paper proposes the DD-a-OFDM scheme for high-mobility communications in 6G and beyond. First, we present a comprehensive system model for DD-a-OFDM, which preserves the conventional TF domain resource multiplexing structure while introducing novel DD domain channel estimation and equalization techniques. Second, we rigorously prove that TF domain ICI can be characterized as Gaussian interference for CSF estimation purposes and derive the corresponding CRLBs. Third, we design the ML-based CSF estimation algorithm for the DD-a-OFDM system. Extensive simulation results validate our design, demonstrating that DD-a-OFDM achieves superior BER performance compared to conventional OFDM systems through accurate LTV channel parameter estimation. Furthermore, the proposed scheme {\color{black}can achieves theoretical estimation limits equivalent to EP-OTFS systems under identical pilot overhead and reduce} pilot overhead relative to EP-OTFS systems by employing a discrete TF domain pilot-based CSF estimation framework. These findings underscore the potential of DD domain signal processing to enhance commercial OFDM systems in high-mobility environments.\par

Several research directions merit further investigation. First, the optimal utilization of channel diversity in DD-a-OFDM based on CSF information requires deeper exploration. While advanced equalizers can potentially exploit this diversity, developing solutions that maintain the low complexity of OFDM while achieving OTFS-comparable performance remains crucial. Second, the impact of residual ICI on CSF estimation accuracy needs to be addressed. Iterative techniques using detected symbols to mitigate ICI effects could yield significant reductions in estimation MSE. These improvements would render DD-a-OFDM a more compelling solution for high-mobility communications.

\vspace{12pt}
\end{document}